\documentclass[twocolumn,showpacs,preprintnumbers,amsmath,amssymb,unsortedaddress]{revtex4}

\usepackage{graphicx}
\usepackage{dcolumn}
\usepackage{bm}

\usepackage{color}

\def \bB {{\bf B}}

\def \bE {{\bf E}}

\def \bj {{\bf j}}

\def \bv {{\bf v}}
\def \bw {{\bf w}}

\def \del {{\bm \nabla}}
\def \grad {{\bm \nabla}}
\def \half {\textstyle{\frac{1}{2}}}
\def \div {{\bm \del} \cdot}
\def \curl {{\bm \del} \times}
\def \p {\partial}

\def \dB {\partial{\bB}}
\def \dt {\partial{t}}

\def \. {\cdot}
\def \x {\times}

\begin{document}
  

\title{3D Null Point Reconnection Regimes}

\author{E.~R.~Priest} 
\affiliation{Mathematics Institute, St Andrews University, St Andrews KY16 9SS, UK}
\author{D.~I.~Pontin}
\affiliation{Division of Mathematics, University of Dundee, Dundee DD1 4HN, UK}

\date{\today}

\begin{abstract}
Recent advances in theory and computational experiments have shown the need to refine the previous categorisation of magnetic reconnection at three-dimensional null points -- points at which the magnetic field vanishes. We propose here a division into three different types, depending on the nature of the flow near the spine and fan of the null. The spine is an isolated field line which approaches the null (or recedes from it), while the fan is a surface of field lines which recede from it (or approach it).  

So-called {\it torsional\ spine\ reconnection} occurs when field lines in the vicinity of the fan rotate, with current becoming concentrated along the spine, so that nearby field lines undergo rotational slippage.  In {\it torsional\ fan\ reconnection} field lines near the spine rotate and create a current that is concentrated in the fan with a rotational flux mismatch and rotational slippage. In both of these regimes, the spine and fan are perpendicular and there is no flux transfer across spine or fan.  The third regime, called {\it spine-fan reconnection}, is the most common in practice and combines elements of the previous spine and fan models. In this case, in response to a generic shearing motion, the null point collapses to form a current sheet that is focused at the null itself, in a sheet that locally spans  both the spine and fan. In this regime the spine and fan are no longer perpendicular and there is flux transfer across both of them.
\end{abstract}


\maketitle

\section{Introduction}\label{sec:1}
Magnetic reconnection is a fundamental process of energy release that lies at the core of many dynamic phenomena in the solar system such as solar flares, coronal heating events, geomagnetic substorms and flux transfer events. Reconnection in three dimensions has been shown to be completely different in many fundamental respects from the classically studied process in two dimensions \citep{schindler88,hesse88,priest03}. The main thrust of reconnection theory at present is to understand the different ways in which it may take place in three dimensions \citep[e.g., the books][]{priest00,birn07}. A key point is that in three dimensions reconnection occurs where a component of the electric field parallel to the magnetic field is present -- and this can be in many different field configurations. For example, reconnection may occur either at null points \citep[e.g.,][]{craig95,priest96a,craig98,craig00,hornig03,heerikhuisen04,pontin05a,pontin06a} or in the absence of null points at quasi-separatrix layers or hyperbolic flux tubes \citep{priest95a,demoulin96a,demoulin96b,demoulin97d,hornig98,titov03a,hesse05,demoulin06a,titov07a,titov09} or it may occur along separators that join one null point to another \citep{priest96a,longcope96a,galsgaard96b,galsgaard00b,longcope01,parnell04,priest05a,longcope05,parnell08a}. 

Null points are common in the solar atmosphere \citep{filippov99,schrijver02,longcope03,close04c} and are sometimes implicated in solar flares and coronal mass ejections \citep{longcope96b,fletcher01,aulanier00,aulanier05a,aulanier06a,cook09}. Three-dimensional collapse of a null has been described \citep{bulanov84,bulanov97,klapper96,parnell96,parnell97} and stationary resistive flows near them have been modelled \citep{craig96,craig99,titov00}. In particular, for a linear null and uniform magnetic diffusivity, \citet{titov00} discovered field-aligned flows when the spine current is small and spiral field-crossing flows which do not cross the spine or fan when the spine current exceeds a critical value.

A three-dimensional null point possesses two different classes of field lines that connect to the null: for a so-called {\it positive null point}, a surface of field lines (called a {\it fan} by \citet{priest96a}) recede from the null, while an isolated field line (called the {\it spine} of the null) approaches  it from two directions; for a {\it negative null point}, on the other hand the fan approaches the null, while the spine recedes from it. (For an alternative nomenclature see Ref.~\cite{lau90}.) The different types of linear null were categorised by \citet{parnell96}. The generic null in a potential magnetic field is an improper radial null, with the fan perpendicular to the spine and the field lines in the fan approaching or receding from essentially two directions (Fig.~\ref{fig:1}b). A particular case is the proper radial null in which the field lines in the fan are radial (Fig.~\ref{fig:1}a). The effect of a current along the fan is to make the fan and spine no longer perpendicular (Fig.~\ref{fig:3}b), whereas a strong enough current along the spine makes the fan field lines spiral (Fig.~\ref{fig:3}a).

There have been three steps towards categorising reconnection at a null point due to (i) analytical ideal modelling, (ii) kinematic resistive modelling and (iii) computational experiments.  The initial analytical ideal treatment by \citet{priest96a} aimed to understand the types of ideal motions that are possible in the environment of a null point. They supposed that the nature of reconnection is determined to a large extent by the nature of the large-scale flows: they suggested that an ideal flow across the fan would drive {\it spine\ reconnection}, in which a current forms along the spine, whereas an ideal flow across the spine would drive {\it fan\ reconnection} with a strong current in the fan. They also proposed {\it separator\ reconnection} with a strong current along a separator joining two nulls.  

Since then, as we shall see in this paper, although behaviour reminiscent of the early spine and fan models may be observed in certain limiting situations, recent numerical experiments have suggested different forms of spine and fan reconnection and also a hybrid spine-fan regime as being the generic modes that occur in practice. However, the existence of separator reconnection has been well confirmed by a series of numerical experiments \citep{galsgaard96b,galsgaard00a,parnell04,parnell08a} and its importance in the solar corona has been stressed \citep{longcope01,longcope05}.  In addition, quasi-separatrix layer reconnection (called slip-running reconnection by \citet{aulanier06a}) has been confirmed in numerical experiments \cite{birn00,pontin05c,aulanier05a,mellor05,demoortel06a,demoortel06b,wilmot07a} and in bright point and flare simulations \citep{demoulin93b,demoulin97a,demoulin97d,masson09a,torok09}.

Our aim here is simply to look more closely at the nature of reconnection at a 3D null point and to propose a new categorisation to replace spine reconnection and fan reconnection.  In the next section it is necessary to summarise the main results from  theory and computational experiments on null-point reconnection and to reinterpret them in the light of the new regimes of reconnection that we are proposing. In the following sections we consider in turn the properties of the three new types of reconnection, namely, {\it torsional\ spine\ reconnection}, {\it torsional\ fan\ reconnection} and the most common regime {\it spine-fan\ reconnection}.

\section{Theory and numerical experiments} \label{sec:2}
\subsection{Null Points}\label{sec:2A}
The simplest linear null point (for which the magnetic field increases linearly from the null) has field components
\begin{equation}
(B_{x},B_{y},B_{z})=\frac{B_{0}}{L_{0}}(x,y,-2z)
\label{eq:1}
\end{equation}
in Cartesian coordinates or
\begin{displaymath}
(B_{R},B_{\phi},B_{z})=\frac{B_{0}}{L_{0}}(R,0,-2z)
\end{displaymath}
in cylindrical polars, so that $\div \bB=0$ identically, where $B_{0}$ and $L_{0}$ are constant.  The field lines are given by 
\begin{displaymath}
y=cx,\ \ \ \ \ \ \ \ \ z=k/x^{2},
\end{displaymath}
where $c$ and $k$ are constants. The $z$-axis is the spine and the $xy$-plane is the fan. 

\begin{figure}
\centering
\includegraphics[height=.25\textheight]{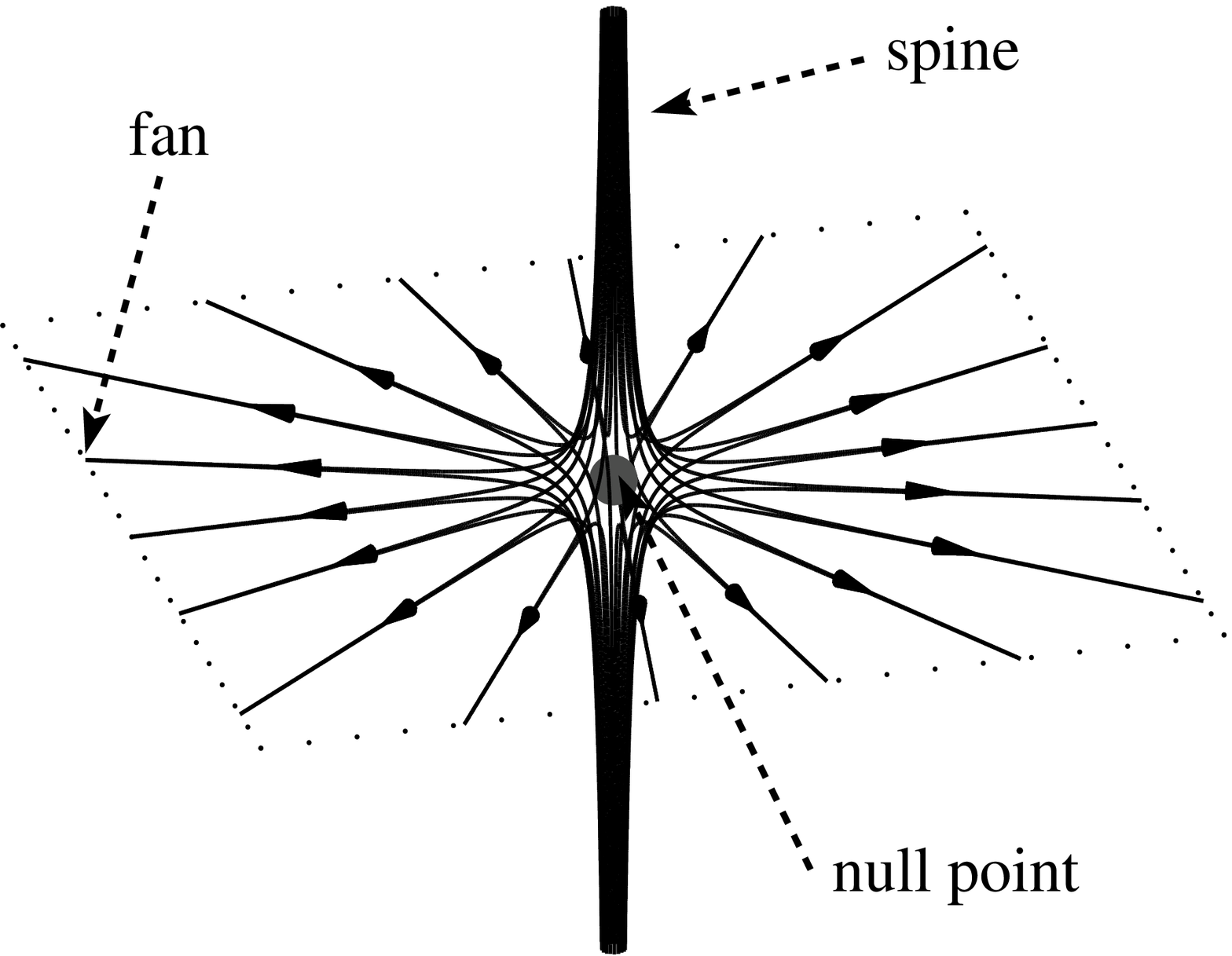}
\includegraphics*[height=.25\textheight]{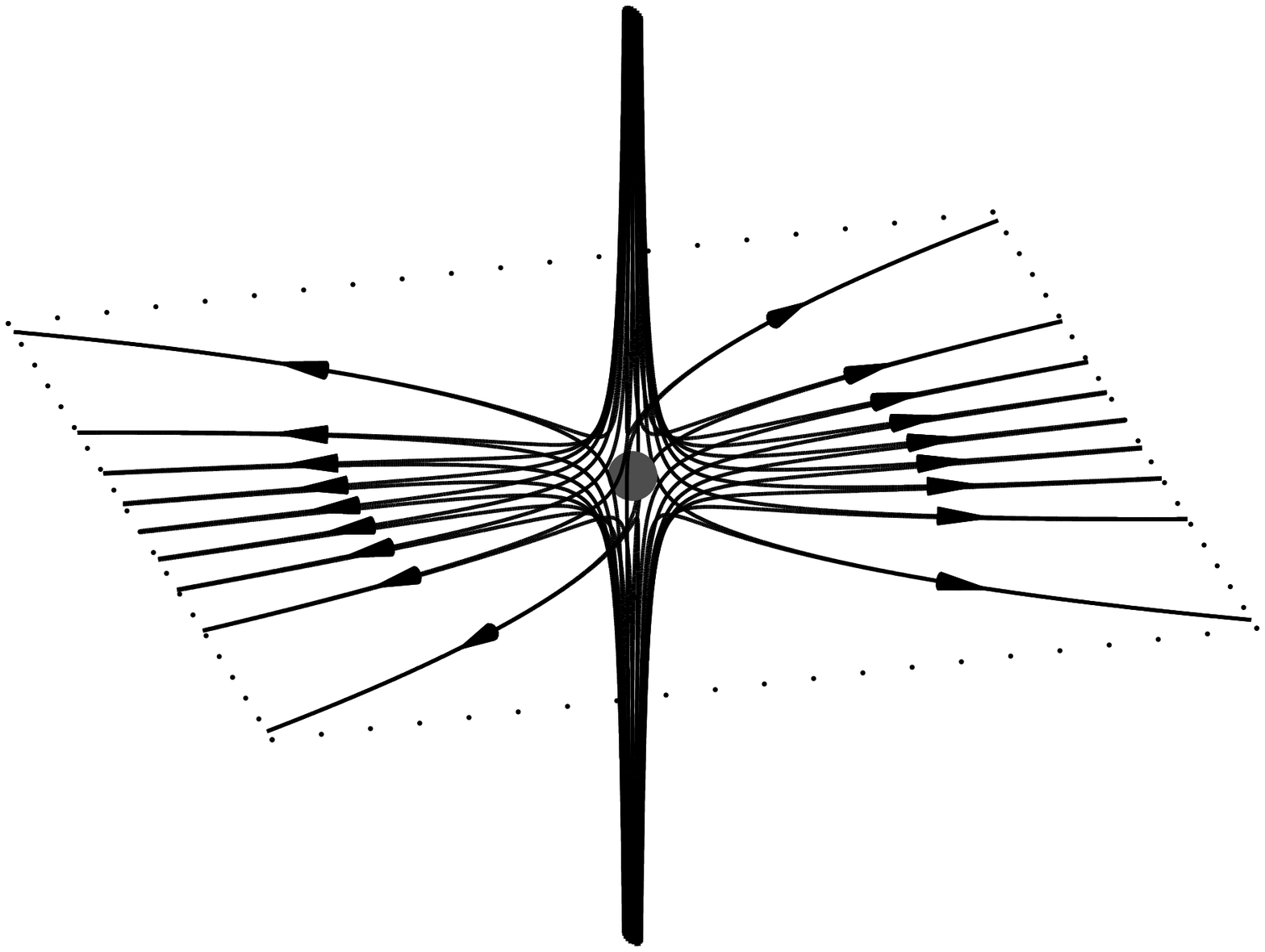}
\caption{Field lines for (a) a proper radial null and (b) an improper radial null.}
\label{fig:1}
\end{figure}

For this so-called {\it proper\ radial\ null} the fan field lines are straight (Figure \ref{fig:1}a). 
It is a particular member (with $a=1$) of a wider class of current-free improper radial null points ($a\neq 1$) with curved fan field lines, having field components
\begin{displaymath}
(B_{x},B_{y},B_{z})=\frac{B_{0}}{L_{0}}[x,ay,-(a+1)z].
\end{displaymath}
This is the generic form for a current-free null, since the proper radial null is structurally unstable in the sense that it occurs only for a particular value of $a$, but for simplicity much of the theory so far has used a proper radial null.

More generally, each of the three field components of a linear null may be written in terms of three constants, making nine in all. However, \citet{parnell96} built on earlier work \citep{cowley73b,fukao75,greene88} and showed, by using $\div \bB=0$, by normalising and by rotating the axes, that the nine constants may be reduced to four constants $(a,b,j_{\parallel},j_{\perp})$ such that
\begin{gather*}
\begin{pmatrix}B_{x} \\ B_{y}\\ B_{z}\end{pmatrix}=
\frac{B_{0}}{L_{0}}\begin{pmatrix}1 & \half(b-j_{\parallel}) & 0\\ 
\half(b+j_{\parallel}) & a & 0\\ 
0 & j_{\perp} & -a-1
\end{pmatrix}
\begin{pmatrix}x \\ y\\ z\end{pmatrix},
\end{gather*}
where $ j_{\parallel}/\mu$ is the current parallel to the spine and 
$j_{\perp}/\mu$ is the current perpendicular to the spine.  Furthermore, both nulls and separators are susceptible to collapse to form current sheets when the boundary conditions allow it \citep{bulanov84,longcope96a,pontin05a}.

\subsection{Kinematic Ideal Models}\label{sec:2B}
The effects in the ideal region around a 3D null of steady reconnection were studied in the kinematic regime by \citet{priest96a} extending earlier ideas \cite{lau90}. They solved the equations
\begin{eqnarray}
\bE+\bv \x \bB={\bf 0}
\label{eq:2}
\end{eqnarray}
and 
\begin{eqnarray}
\curl \bE = {\bf 0}
\label{eq:3}
\end{eqnarray}
for $\bv$ and $\bE$ when $\bB$ is given by Equation (\ref{eq:1}) and a variety of different boundary conditions are imposed.  

In particular, Eq.~(\ref{eq:3}) implies that $\bE=\del \Phi$ and then the component of Equation (\ref{eq:2}) perpendicular to $\bB$ yields
\begin{eqnarray}
\bB \. \del \Phi=0,
\label{eq:4}
\end{eqnarray}
which, for certain imposed boundary conditions, may be integrated along field lines (characteristics) to determine the value of $\Phi$ (and therefore $\bE$) throughout the volume.  Then the component of Eq. (\ref{eq:2}) perpendicular to $\bB$ determines the plasma velocity normal to $\bB$ everywhere as
\begin{eqnarray}
\bv_{\perp}=\frac{\del \Phi \x \bB}{B^{2}}.
\label{eq:5}
\end{eqnarray}

\begin{figure}
\centering
\includegraphics[height=.30\textheight]{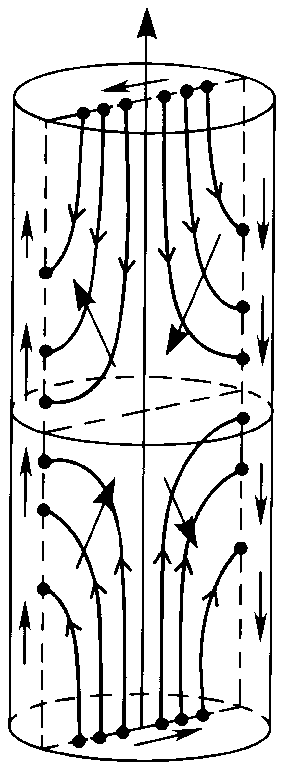}
\includegraphics[height=.30\textheight]{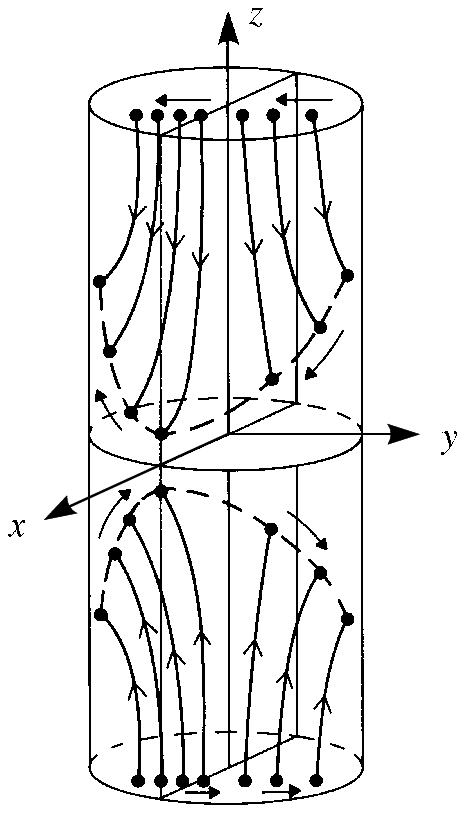}
\caption{Regimes envisaged from ideal motions: (a) Spine reconnection with a strong spine current driven by continuous motions across the fan. (b) Fan reconnection with a strong fan current and flipping of field lines above and below the fan produced by continuous motions across the spine.}
 \label{fig:2}
\end{figure}
If a continuous flow is imposed across the fan (Figure \ref{fig:2}a), singularities in $\bE$ and $\bv$ are produced at the spine. \citet{priest96a} speculated that this would produce a strong current at the spine in what they dubbed {\it spine\ reconnection}. They considered the effect of diffusion in a preliminary manner, but they were unable at the time to resolve the singularities at the spine. As an example, they considered flows with no $\phi$-component and an electric field of the form $E_{\phi}=v_{e}B_{0} \sin \phi$ giving rise to a velocity
\begin{eqnarray}v_{\perp R}=\frac{2E_{\phi}L_{0}^{2}z/B_{0}}{R(R^{2}+4z^{2})},\ \ \ \ \ \  \ \ \  v_{\perp z}=\frac{E_{\phi}L_{0}^{2}z/B_{0}}{R^{2}+4z^{2}}, 
\nonumber
\end{eqnarray}
for which $v_{\perp z}$ is continuous at the fan $z=0$, while $v_{\perp R}$ is singular at the spine $R=0$.

If, on the other hand, a continuous flow is imposed across the spine (Figure \ref{fig:2}b), singularities are produced at the fan together with a strong {\it flipping\ flow} (that \citet{priest92b} had previously discovered). \citet{priest96a} suggested that this would produce a strong current at the fan in what they dubbed {\it fan\ reconnection}. A particular example is given in terms of ${\bar x}=x/L$, ${\bar y}=y/L$, ${\bar z}=z/L$ by a potential of the form $\Phi=v_{e}B_{e}[{\bar x}^{2}{\bar z}/(4+{\bar y}^{2}{\bar z)}^{\half}]$, which produces a flow field
 \begin{eqnarray}
 (v_{\perp {\bar x}},v_{\perp {\bar y}},v_{\perp {\bar z}})=\frac{v_{e}}{({\bar x}^{2}+{\bar y}^{2}+4{\bar z}^{2})(4+{\bar y}^{2}{\bar z})^{3/2}}\times\ \ \ \ \ \ \ \ \ \ 
 \nonumber\\  
\left(\frac{2{\bar x}{\bar y}{\bar z}({\bar z}^{3}-1)}{{\bar z}^{1/2}},\frac{2({\bar x}^{2}+4{\bar z}^{2}+{\bar y}^{2}{\bar z}^{3})}{{\bar z}^{1/2}},(4+{\bar y}^{2}{\bar z}+{\bar x}^{2}{\bar z}){\bar y}{\bar z}^{\half}\right),
\nonumber
 \end{eqnarray}
for which $v_{\perp {\bar y}}$ is continuous on the planes ${\bar z}=\pm 1$, while $v_{\perp {\bar x}}$ and $v_{\perp {\bar y}}$ are singular at the fan (${\bar z}=0$).  However, this analysis left open the questions as to whether it is possible to resolve the singularity and also whether these pure states are likely to be set up in practice.

\subsection{Kinematic Resistive Models}\label{sec:2C}
The next step in the theory was to consider the effect in 3D of an isolated diffusion region where frozen-in flux breaks down and the induction equation is typically of the form
 \begin{eqnarray}
 \frac{\p \bB}{\p t}=\curl (\bv \x \bB)+\eta \nabla^{2}\bB.
\nonumber
\end{eqnarray}
Reconnection in 3D is very different in many respects from that in 2D. 

In 2D, a differentiable flux-transporting velocity $\bw$ \citep{hornig96} satisfying
 \begin{eqnarray}
\frac{\dB}{\dt}=\curl (\bw \x \bB)
\nonumber
\end{eqnarray} 
always exists apart from at the X-point itself. This velocity has a hyperbolic singularity at an X-type null point, where the reconnection takes place. The magnetic flux moves at the velocity $\bw$ and slips through the plasma, which itself moves at $\bv$. Furthermore, the mapping of the field lines in 2D is discontinuous at the separatrix field lines that thread the X-point. This mapping discontinuity is associated with the fact that field lines break and reconnect at one point, namely, the X-point.  While they are in the diffusion region, field lines preserve their connections everywhere, except at the X-point. Two flux tubes that move into the diffusion region break and rejoin perfectly to form two new flux tubes that move out.  

In 3D, surprisingly, none of the above properties carry over and so the nature of reconnection is profoundly different \citep{priest03}.  First of all, a single flux tube velocity ($\bw$) does not generally exist \citep{hornig96,hornig01} since $\bE \. \bB \neq 0$, but it may be replaced by a pair of flux velocities describing separately what happens to field lines that enter or leave the diffusion region  \citep{priest03}. Secondly, the mapping of field lines is continuous if there is no 3D null point or separatrix surface.  Thirdly, as they move through a 3D diffusion region, magnetic field lines continually change their connections.  Fourthly, two tubes don't generally break and reform perfectly to give two other flux tubes: rather, when the two flux tubes are partly in the diffusion region and so are in the process of reconnecting, they split into four parts, each of which flips in a different manner, a manifestation of the continual change of connections. (Note that in general in 2D and 3D the flux velocity ${\bf w}$ is non-unique. We choose here to consider the case in which we select $\bw$ by insisting that $\bw=\bv$ in the ideal region. The crucial distinction is that in 2D a single $\bw$ exists (and is singular), while in 3D reconnection {\it no single} velocity $\bw$ exists that satisfies (\ref{eq:9}) together with the constraint that $\bw=\bv$ in the ideal region. See Refs.~\citep{hornig96,hornig01} for further discussion.)

The first attempt to model kinematically the effect of an isolated diffusion region was by \citet{hornig03} who set up a formalism and applied it to a case without null points. 
They solved
\begin{eqnarray}
\bE+\bv \x \bB=\eta\ \bj,
\label{eq:6}
\end{eqnarray}
where $\curl \bE = {\bf 0}$, $\bj=\curl \bB/\mu$ and $\div \bB=0$.
The idea was to impose a sufficiently simple magnetic field that both the mapping and the inverse mapping of the field can be found analytically. Then, after writing $\bE=\del \Phi$, the integral of the component of (\ref{eq:6}) parallel to $\bB$ determines $\Phi$ everywhere as an integral
\begin{eqnarray}
\Phi =\int \frac{\eta\ \bj \. \bB}{B}\ ds +\Phi_{e}
\nonumber
\end{eqnarray}
along field lines, in terms of the values ($\Phi_{e}$) at one end of the field lines and the distance $s$ along field lines. More simply in terms of a dimensionless stretched distance $S$ such that $ds/B=L_{0}dS/B_{0}$,
\begin{equation}
\Phi =\int \frac{\eta\ L_{0}\  \bj \. \bB}{B_{0}}\ dS +\Phi_{e}.
\label{eq:7}
\end{equation}

One way of isolating the reconnection region in these kinematic solutions is by choosing a form of $\eta$ that is localised. So-called {\it pure} solutions have $\Phi_{e}\equiv 0$ and produce counter-rotating (or flipping) flows of field lines that link the diffusion region. The rate of flux reconnection is calculated by evaluating the integral
\begin{eqnarray}
\frac{d\Phi_{mag}}{dt} =\int E_{\parallel} ds
\label{eq:8}
\end{eqnarray}
along a field line through the diffusion region \citep{schindler91,hesse05}. Then the flow normal to the field lines is determined by the component of Equation (\ref{eq:6}) perpendicular to $\bB$ as
\begin{eqnarray}
\bv_{\perp}=\frac{(\del \Phi -\eta\ \bj) \x \bB}{B^{2}}.
\label{eq:9}
\end{eqnarray}
These solutions may be regarded as either kinematic (i.e., satisfying just the induction equation) or as fully dynamic in the limit of uniform density and slow flow (since they also satisfy the equations $\div \bv=0$ and $\del p=\bj \x \bB$).

\begin{figure}
\centering
\includegraphics[height=.35\textheight]{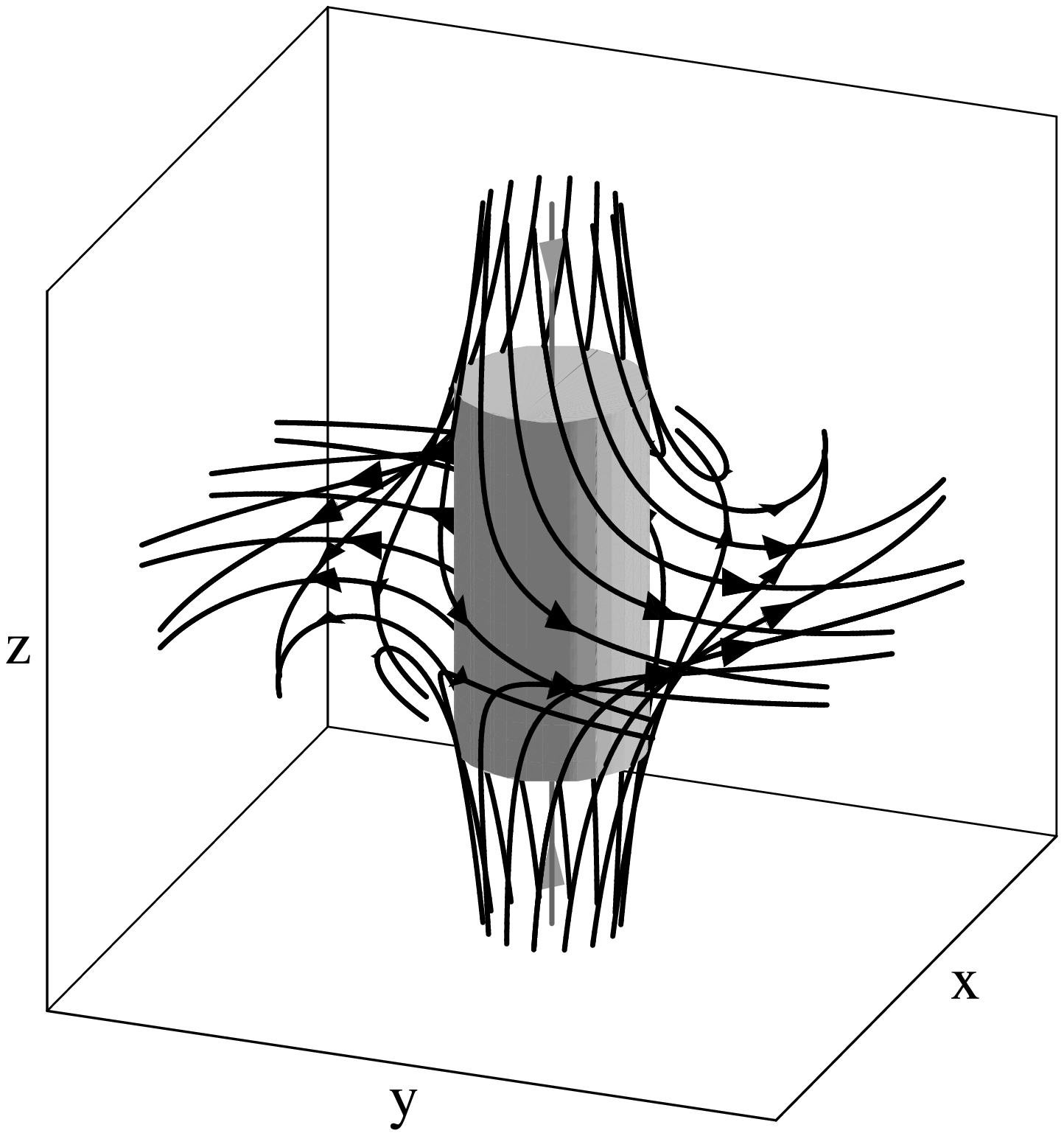}
\includegraphics[height=.35\textheight]{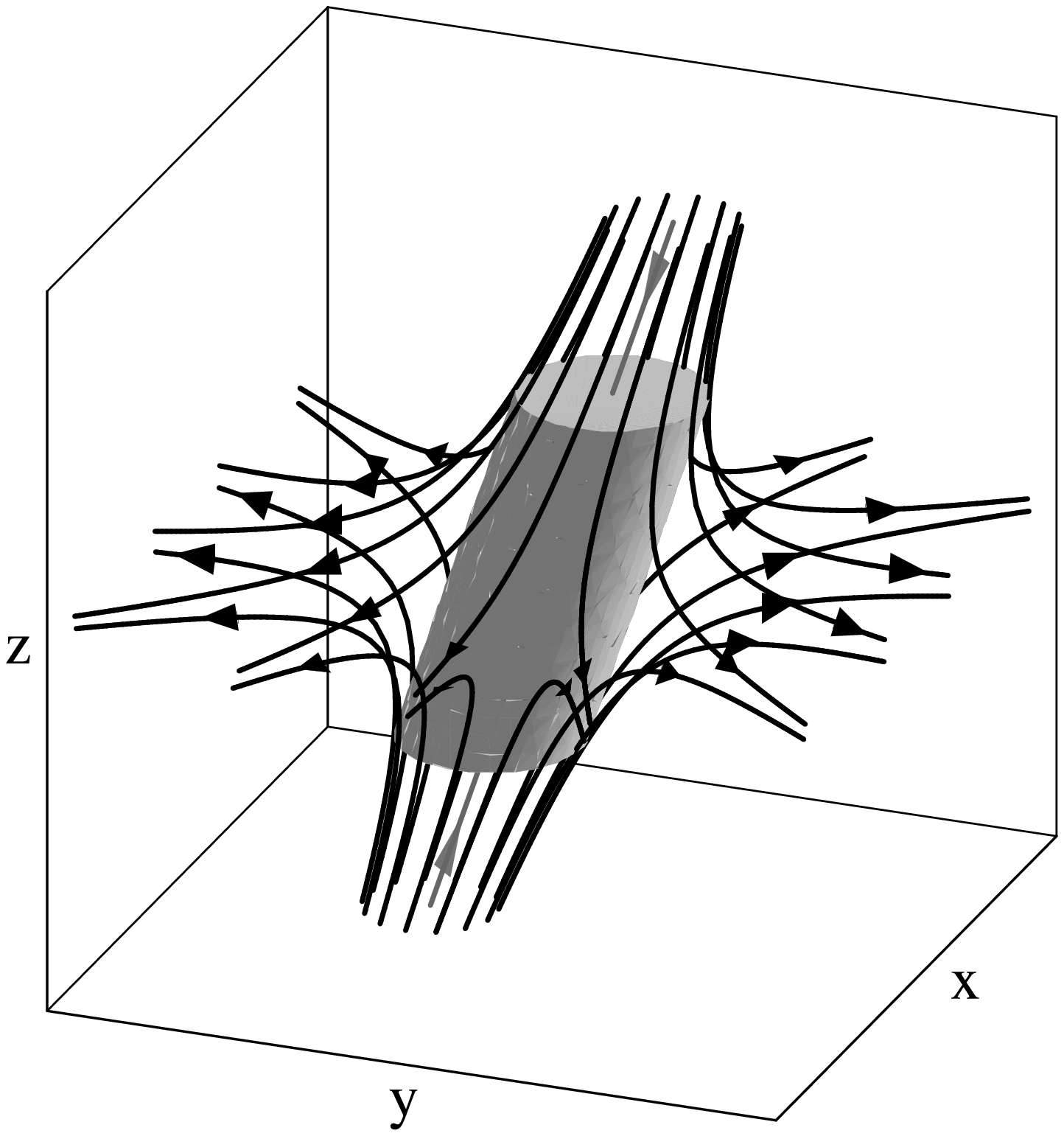}
\caption{The field near a null point with (a) uniform spine current and (b) uniform fan current.}
 \label{fig:3}
\end{figure}
\citet{pontin04a} applied this formalism to determine the behaviour of the magnetic flux when an isolated diffusion region contains a spiral null point, i.e.~a null with current directed parallel to the spine line. The imposed magnetic field was
\begin{eqnarray}
(B_{x},B_{y},B_{z})=\frac{B_{0}}{L_{0}}\left(x-\half  {\bar j_{0}} y,y+\half  {\bar j_{0}} x,-2z\right)
\nonumber
\end{eqnarray}
or
\begin{eqnarray}
(B_{R},B_{\phi},B_{z})=\frac{B_{0}}{L_{0}}\left(R,\half  {\bar j_{0}} R,-2z\right)
\label{eq:10}
\end{eqnarray}
in cylindrical polars, with the spine and current both directed along the $z$-axis, where ${\bar j_{0}}$ is a dimensionless current density. The diffusion region was assumed to be a cylinder of radius $a$ and height $2b$ (Figure \ref{fig:3}a).

First of all, a pure elementary solution which describes the core of the reconnection process was obtained by setting the flow to zero outside the volume defined by the `envelope' ($F$) of flux  that threads the diffusion region. Inside $F$ the flow and flux velocities are purely rotational 
(i.e., in the $\phi$-direction), so that there is no flow across either the spine or the fan. The reconnection rate is $\int E_{\parallel}dl$ along the spine, and measures the rate of rotational mis-matching of the flux velocities of field lines entering and leaving the diffusion region.

To this solution any ideal solution ($\Phi_{id}$) may be added and in particular they considered a stagnation-point flow of the form $\Phi_{id}=\phi_{0} x_{0}y_{0}$, which brings flux into $F$ and carries it out again. The result is a transition from O-type to X-type flow near the null when $\phi_{0}$ exceeds a critical value. What this solution suggests, therefore, is that a type of spine reconnection with strong current along the spine direction is possible when there are twisting flows about the spine.  This is quite different from the spine reconnection that was envisaged in \citet{priest96a} and so here we propose to call it  {\it torsional\ spine\ reconnection} and discuss its properties further in Section \ref{sec:3}.

Next, \citet{pontin05b} applied the same approach to a diffusion region ($D$) containing a null point having a uniform fan-aligned current ($B_{0}{\bar j_{0}}/(\mu L_{0})$) in the $x$-direction and field components 
\begin{eqnarray}
(B_{x},B_{y},B_{z})=\frac{B_{0}}{L_{0}}(x,y-{\bar j_{0}}z,-2z).
\nonumber
\end{eqnarray}
The diffusion region was assumed to have the shape of a disc of radius $a$ and height $2b$ (see Fig.~\ref{fig:3}(b)), inside which the magnetic diffusivity decreases smoothly and monotonically from the null to zero at its boundary. Outside $D$ it vanishes.  

The resulting plasma flow was surprisingly found to be quite different from the fan reconnection of \citet{priest96a}, since it is found to cross both the spine and fan of the null.  Field lines traced from footpoints anchored in the fan-crossing flow are found to flip up and down the spine, whereas those that are traced from the top and bottom of the domain flip around the spine in the fan plane, as envisaged by \citet{priest96a}.  The reconnection rate is again given by an integral of the form (\ref{eq:8}), this time along the fan field line parallel to the direction of current flow (here the $x$-axis). For such a mode of reconnection this expression can be shown to coincide with the rate of flux transport across the fan (separatrix) surface \cite{pontin05b}.



It is possible to find a solution that has similar field line behaviour to the pure fan reconnection envisaged by \citet{priest96a}, with  flow across the spine but not the fan, by adopting instead a field of the form $(B_{0}/L_{0})(x,y-{\bar j_{0}}z^{3}/L_{0}^{2},-2z)$ 
with a fan $x$-current $3B_{0}{\bar j_{0}}z^{2}/(\mu L_{0}^{3})$ 
(see Ref.~\cite{pontin05b}). It is also possible to model pure spine reconnection with  flow across the fan but not the spine by considering $(B_{0}/L_{0})(x,y,{\bar j_{0}}y^{3}/L_{0}^{2}-2z)$.
with a fan $x$-current $3B_{0}{\bar j_{0}}y^{2}/(\mu L_{0}^{3})$. 
Both of these fields have a vanishing current at the null. 
However, a key property of a null point is the hyperbolic field structure, which tends to focus disturbances and thus generate non-zero currents at the null for the {\it primary\ reconnection\ modes}. The above pure spine and fan solutions should therefore not  be considered as fundamental or primary reconnection modes but as {\it secondary\ reconnection\ modes} in the sense that the current vanishes at the null. 

It has been suggested that solutions for spine reconnection in incompressible plasmas \cite{craig96} may not be dynamically accessible, and while incompressible fan solutions \cite{craig95} are dynamically accessible \cite{craig98, pontin07c, titov04, tassi05}, this breaks down when the incompressibility assumption is relaxed \cite{pontin07c}. 
It turns out that the generic null point reconnection mode that is observed in numerical experiments in response to shearing motions is one  in which there is a strong fan current with flow across both spine and fan, and which is in some sense a combination of the spine and fan reconnection of \citet{priest96a}. We propose here to call it  {\it spine-fan\ reconnection} and discuss its properties further in Section \ref{sec:5}.

\subsection{Numerical Experiments}\label{sec:2D}
Several numerical experiments have been conducted in order to go beyond the constraints of analytical theory and to shed more light on the nature of reconnection at a 3D null. The aim was also to see whether the types of reconnection envisaged qualitatively could indeed take place in practice and to discover whether any other regimes are possible.

First of all, \citet{galsgaard03a} investigated propagation of a helical Alfv{\'e}n wave towards the fan plane, launched by a rotational driving of the field lines around  the spine. This led to the concentration of current in the fan plane and suggests the possibility of {\it torsional\ fan\ reconnection} which we shall propose in Section \ref{sec:4}. (For highly impulsive driving coupling to a fast mode wave that wraps around the null was also observed.) On the other hand \citet{pontin07a} used a resistive MHD code to show how rotational disturbances of field lines in the vicinity of the fan plane can also produce the strong currents along the spine that are symptomatic of {\it torsional\ spine\ reconnection} (Section \ref{sec:3}).

Then \citet{pontin05a} used an ideal Lagrangian relaxation code to follow the formation of current sheets by the collapse of a line-tied 3D null in a compressible plasma. This was a result of the focussing of externally generated large-scale stresses in the field in response to an initial shearing of either the spine axis or fan plane. Building on a previous linear theory by \citet{rickard96}, they found that locally the fan and spine collapse towards each other to form a current sheet singularity.
\begin{figure}
\centering
\includegraphics[height=.29\textheight]{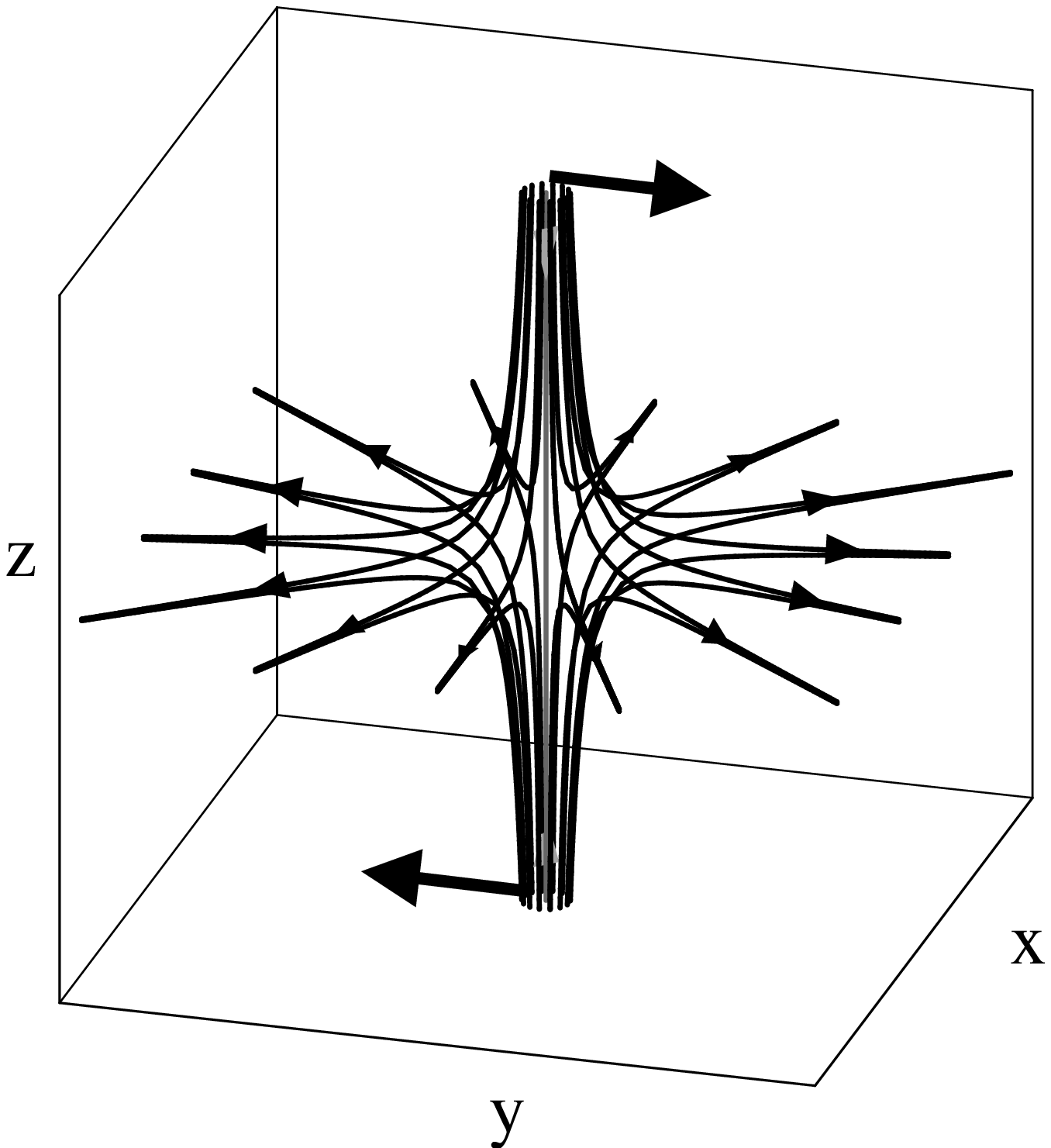}
\includegraphics*[height=.26\textheight]{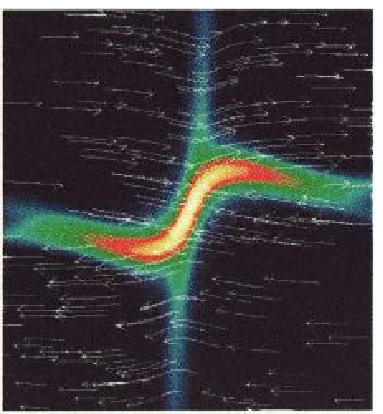}
\caption{(Color) (a) A shearing motion of a spine that is situated on the $z$-axis. (b) The resulting collapse of spine and fan to form {\it spine-fan\ reconnection}, showing the current-density contours (colour) and flow velocity (white) in the $x=0$ plane \citep[After][]{pontin07b}.}
\label{fig:4}
\end{figure}
This was followed up by \citet{pontin07b}, who used a resistive MHD code to investigate the formation and dissipation of the current sheet in response to shearing of the spine, as shown in Figure \ref{fig:4}. The results support the idea of {\it spine-fan\ reconnection} in which current concentrates around the null (in a sheet spanning the spine and fan). Including compressibility does not affect the results qualitatively, except that in the incompressible limit the spine-fan current is found to reduce  purely to a fan current \citep{pontin07c} with behaviour closely resembling earlier fan reconnection models \cite{priest96a,craig95}.  So pure {\it fan\ reconnection} can be either an incompressible limit of {\it spine-fan\ reconnection} or, as we have seen in Section \ref{sec:2C}, the result of a secondary fan current which vanishes at the null.

\section{Torsional spine reconnection}\label{sec:3} 

\begin{figure}
\centering
\includegraphics*[height=.29\textheight]{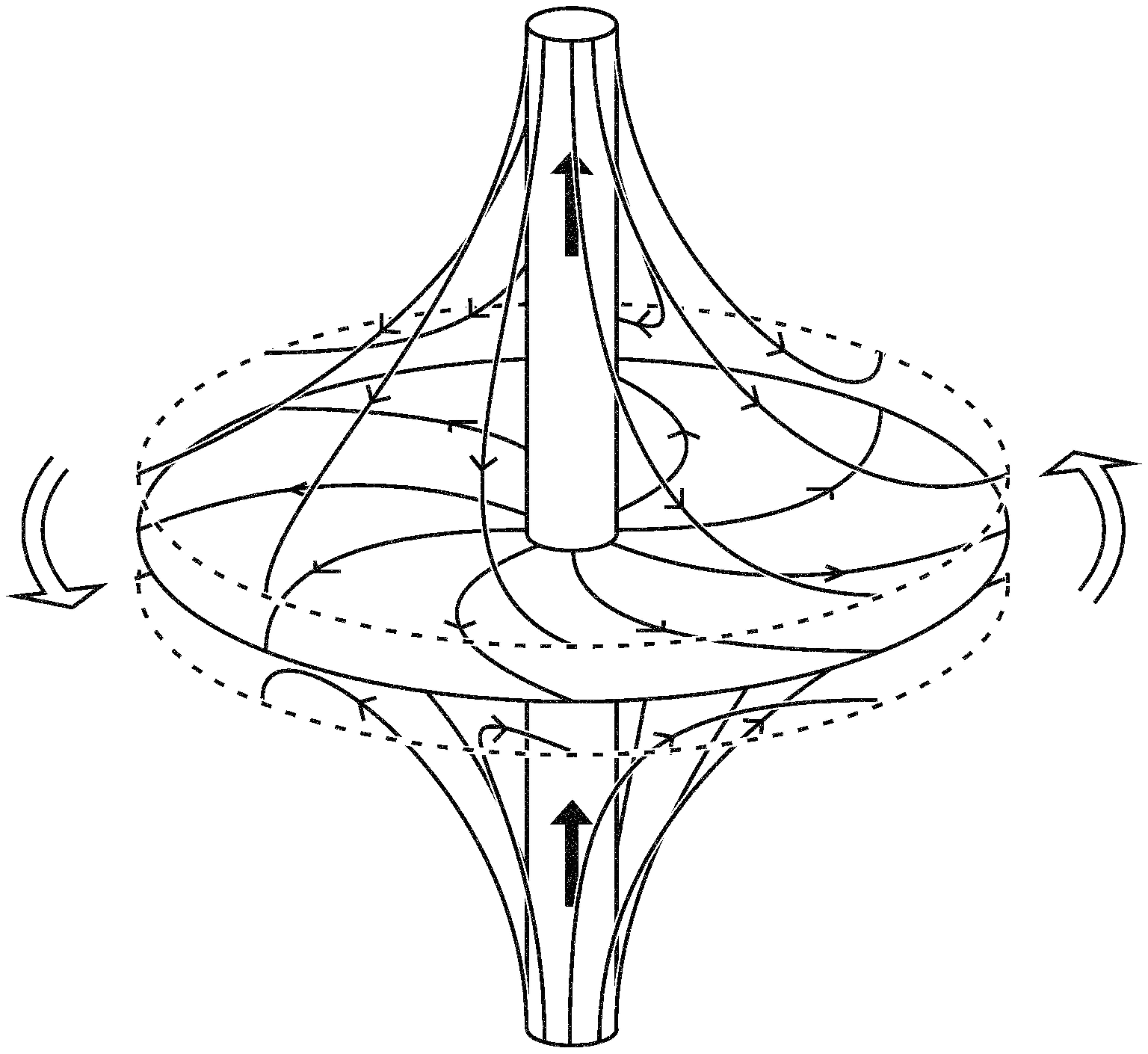}
\includegraphics*[height=.29\textheight]{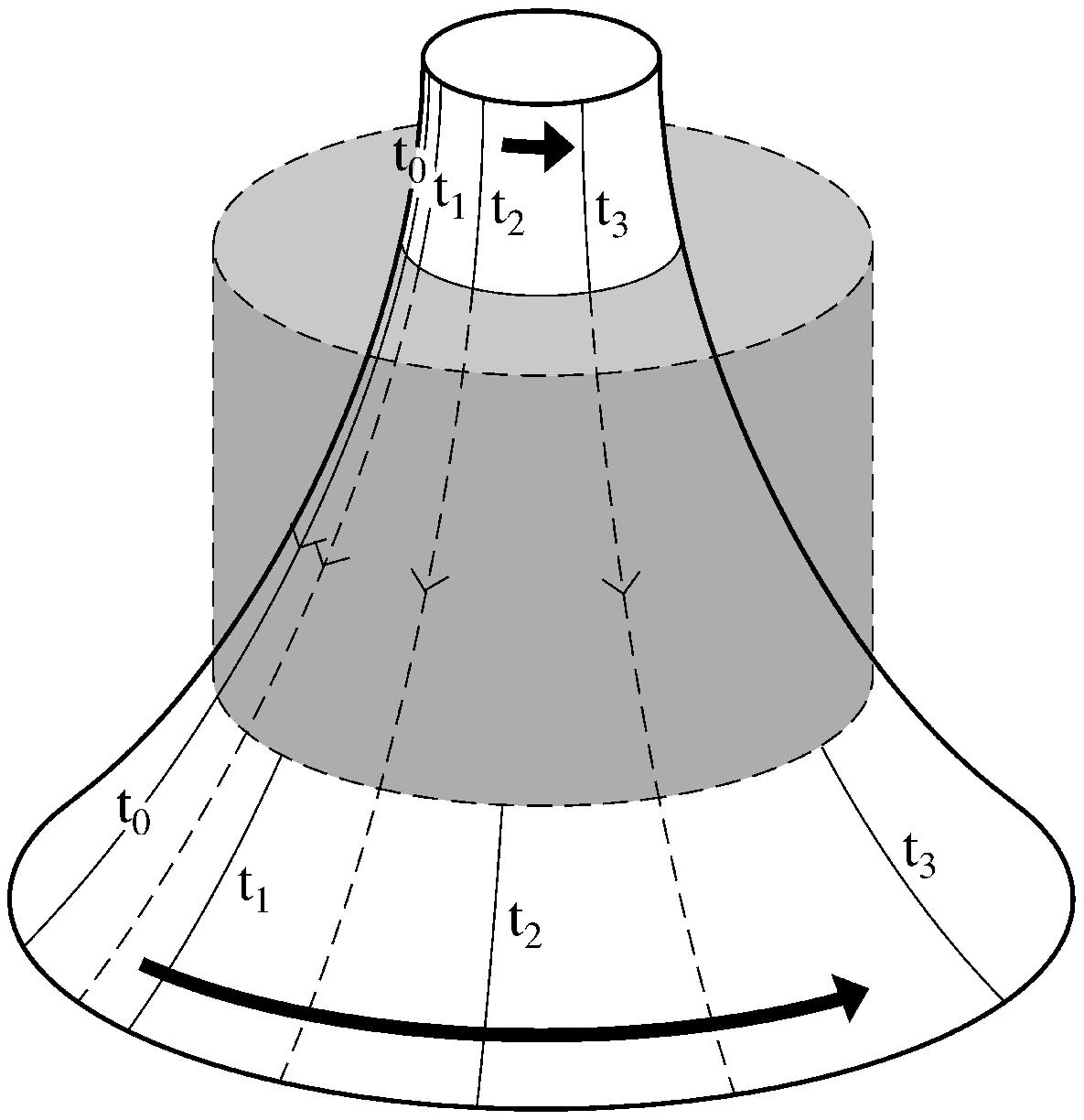}
\includegraphics*[width=.50\textwidth]{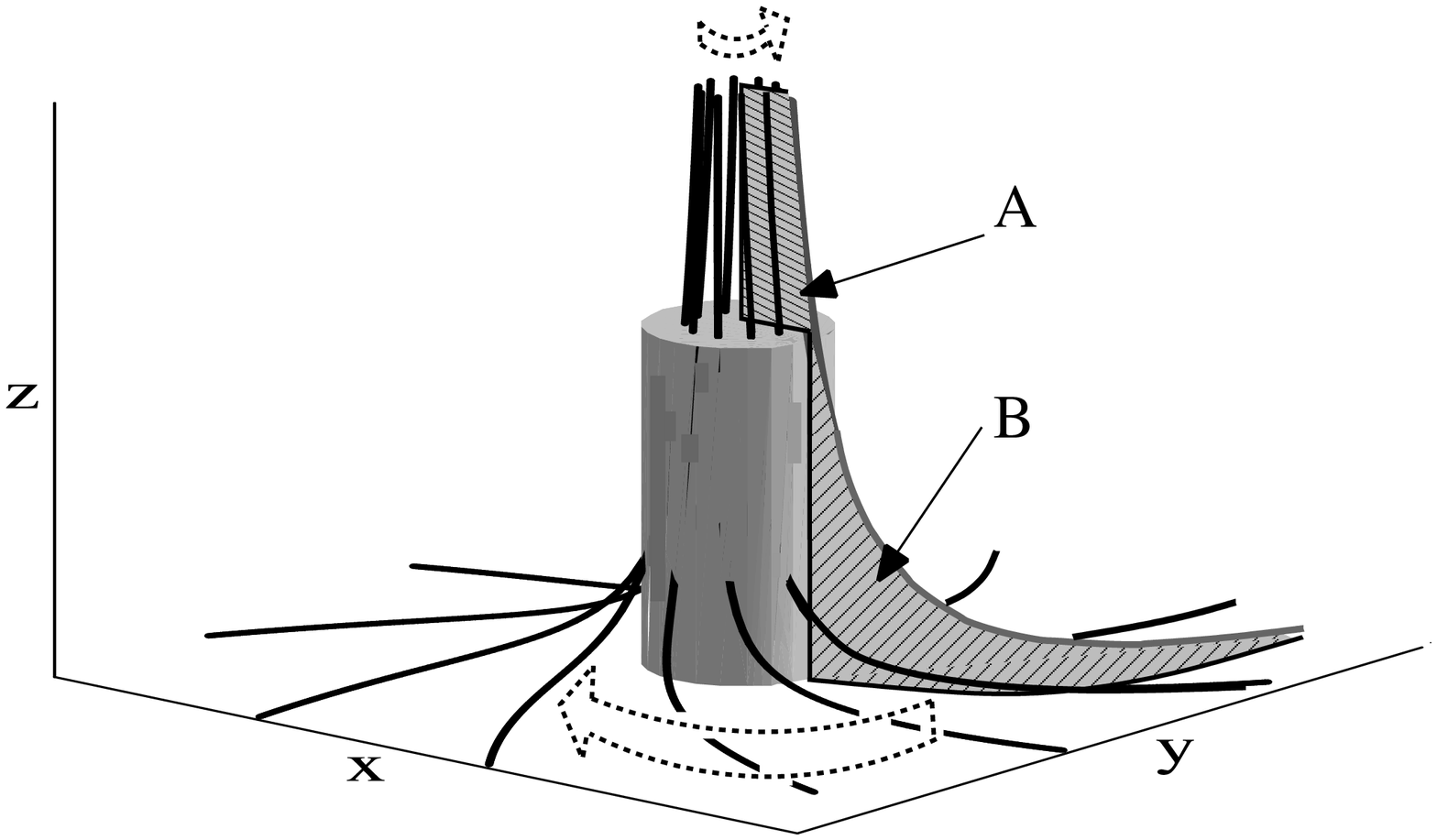}
\caption{(a) A rotational motion of the fan (open arrows) driving torsional spine reconnection with a strong current (solid arrows) along the spine. (b) Rotational slippage of fields entering through the top of the diffusion region on a curved flux surface, showing as solid curves the locations of the plasma elements at $t=t_{1}$, $t=t_{2}$, $t=t_{3}$, that initially ($t=t_{0}$) lay on one field line. (c) The reconnection rate measures a rotational mismatching of flux threading the diffusion region, namely the difference between the rates of flux transport through surfaces A and B.}
\label{fig:5}
\end{figure}

The type of reconnection set up at a 3D null depends crucially on the nature of the flows and boundary conditions that are responsible for the reconnection. Let us suppose first that  a rotation of the fan plane drives a current along the spine and gives rise to torsional spine reconnection, as sketched in Figure \ref{fig:5}a. The nature of the reconnection is that in the core of the spine current  tube there is rotational slippage, with the field lines becoming disconnected and rotating around the spine (see \citet{pontin07a}): Figure \ref{fig:5}b shows on the left side a particular magnetic field line and its plasma elements at $t=t_{0}$; in the upper part of the figure (above the shaded diffusion region) this field line and its attached plasma elements rotate about the spine through positions at times $t_{1}$, $t_{2}$ and $t_{3}$; in the lower part of the figure (below the diffusion region) the plasma elements that were on the field line at $t_{0}$ rotate to positions at $t_{1}$, $t_{2}$ and $t_{3}$ that are on different field lines.
A steady kinematic solution may be found following the approach of Section \ref{sec:2C}. The electric field may be written as the sum ($\bE=\grad \Phi = \del \Phi_{nid} + \del \Phi_{id}$) of a nonideal pure (elementary) solution satisfying
\begin{equation}
\grad \Phi_{nid}+\bv_{nid} \x \bB=\eta \curl \bB,
\nonumber
\end{equation}
and  an ideal solution satisfying
\begin{equation}
\grad \Phi_{id}+\bv_{id} \x \bB={\bf 0}.
\nonumber
\end{equation} 

Consider a spiral null point (Equation \ref{eq:10}) and suppose the diffusion region is a cylinder of radius $a$ and height $2b$ and that  the magnetic diffusivity has the form $\eta = \eta_{0}f(R,z)$, where $f(0,0)=1$ and $f(R,z)$ vanishes on the boundary of  the diffusion region and outside it. 

The field lines for this spiral null  may be obtained by solving
\begin{equation}
\frac{dR}{dS}=\frac{L_{0}B_{R}}{B_{0}}=R, \ \ \ \ R\frac{d\phi}{dS}= {\half}  {\bar j_{0}}R, \ \ \ \ \frac{dz}{dS} = -2 z. \nonumber
\end{equation}
Suppose we start a field line at the point $(R,\phi,z)=(R_{0},\phi_{0},b)$ at $S=0$.  Then the field line equations are
\begin{equation}
R=R_{0}\ e^{S}, \ \ \ \ z=b\ e^{-2S},   \ \ \ \ \phi=\phi_{0}+\half\ {\bar j_{0}}\ S.
\label{eq:11}
\end{equation}
These give a mapping from an initial point $(R_{0},\phi_{0},b)$ to any other point $(R,\phi,z)$ along a field line.  The inverse mapping is 
\begin{equation}
R_{0}=R\ e^{-S}, \ \ \ \phi_{0}=\phi-\half\  {\bar j_{0}}\ S.
\label{eq:12}
\end{equation}
where $S=-{\half}\log(z/b)$.

\subsection{Pure Non-Ideal Solution}\label{sec:3.1}
The pure elementary solution describes the core of the reconnection process. It is obtained following Refs.~\cite{hornig03,pontin04a}  by solving $\bE+\bv \x \bB=\eta\ \bj,$ with $\curl \bE = {\bf 0}$, $\bj=\curl \bB/\mu$ and $\div \bB=0$. Thus we write $\bE=\grad\Phi_{nid}$ with $\Phi_{nid}$ given by Equation (\ref{eq:7}) and set $\Phi_{e}\equiv 0$ so that the flow vanishes outside the diffusion region. Inside the diffusion region the flow and flux velocities have no component across either the spine or the fan.  For the spiral magnetic field $(B_{R},B_{\phi},B_{z})=(B_{0}/L_{0})(R,\half  {\bar j_{0}} R,-2z)$ and the mapping (\ref{eq:11}), $\Phi_{nid}$ becomes
\begin{equation}
\Phi_{nid}\ =\ -\Phi_{nid0}\int \eta/\eta_{0} \ e^{-2S}dS,
\nonumber
\end{equation} 
where $\Phi_{nid0}=2B_{0}b{\bar j_{0}}\eta_{0}/(\mu L_{0})$. Then, once a form for $\eta$ is assumed, this may be integrated to give $\Phi_{nid}(S,R_{0},\phi_{0})$. After using the inverse mapping (\ref{eq:12}), we can then deduce  $\Phi_{nid}(R,\phi,z)$ and therefore $\bE$ and $\bv_{\perp}$ everywhere.  

If a diffusion region is isolated, a change of connectivity of field
lines may be studied, by following field lines anchored in the ideal
region on either side of the diffusion region. A diffusion region is
in general isolated if  $\eta \bj$ is localised in space. In practical
cases in astrophysics, this is likely to be mainly because $\bj$ is
localised but in addition sometimes because as a consequence $\eta$ is also
localised. Some numerical simulations have a localised $\eta$, whereas
others have a uniform $\eta$ or a purely numerical dissipation.  But
the important feature in all these cases is that the product $\eta
\bj$ is localised.  Now, in each of our solutions below, we follow
Refs.~\cite{hornig03,pontin04a,pontin05b} in choosing a spatially
localised $\eta \bj$ by imposing a spatially localised resistivity
profile together with a $\bj$ that is not localised.  The reason for
doing this is to render the mathematical equations tractable, since we
have not yet discovered a way to do so with a localised $\bj$. The
quantitative spatial profiles of physical quantities will depend on
the $\eta$ profile, but the qualitative topological properties of the
field line behaviour in such models are expected to be generic and
independent of the particular profile chosen for $\eta$.  Indeed, the
topological properties of the reconnection models of
Refs.~\cite{hornig03,pontin04a,pontin05b} have been verified by the
numerical simulations \cite{pontin05c,pontin07a,pontin07b}.

There are four regions  with different forms for $\Phi_{nid}$, as illustrated in Figure 6, which shows a vertical cut in the first quadrant of the $Rz$-plane.  In region (1) threaded by field lines that enter the diffusion region (shaded) from above, we assume $\Phi_{nid}(R,z)\equiv 0$, so that there is no electric field or flow. The same is true in region (2) which lies above the flux surface $zR^{2}=ba^{2}$ that touches the upper corner $(a,b)$ of the diffusion region. We calculate below the forms of  $\Phi_{nid}(R,z)$ in the diffusion region (3) and in the region (4) threaded by field lines that leave the diffusion region through its sides.

\begin{figure}
\centering
\includegraphics*[width=.49\textwidth]{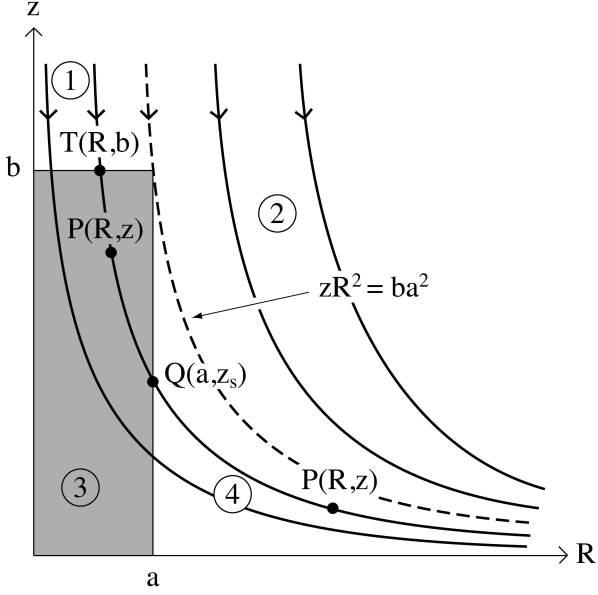}
\caption{The  projection of  magnetic field lines and the diffusion region in the first quadrant of the $R$-$z$ plane, showing 4 different regions (1)-(4) in which $\Phi_{nid}(R,z)$ is calculated.  A magnetic field line whose projection intersects the top of the diffusion region in $T(R,b)$ and the side in $Q(a,z_{s})$  contains typical points $P(R,z)$ inside and beyond the diffusion region. The bounding field line $zR^{2}=ba^{2}$ is shown dashed.}
\label{fig:6}
\end{figure}
For example, let us assume that $\eta$ vanishes outside the diffusion region ($D$) and that inside $D$ it has the form
\begin{equation}
\eta=\eta_{0}\left(1-\frac{R^{4}}{a^{4}}\right)\left(1-\frac{z^{2}}{b^{2}}\right),
\nonumber
\end{equation} 
which peaks at the origin and vanishes on the boundary of $D$. First, we use the mapping (\ref{eq:11}) to substitute for $R$ and $z$, and integrate with respect to $S$ from the point $T(R,b)$ on the top of $D$ to the point $P(R,z)$ inside $D$ (Figure \ref{fig:6}). Then we use the inverse mapping (\ref{eq:12}) to replace $R_{0}$ and $S$, and finally we obtain the potential throughout $D$ (region (3) in Figure \ref{fig:6}) as
\begin{eqnarray}
\Phi_{nid}(R,z)=-{\half}\Phi_{nid0}\left[\left(1-\frac{z}{b}\right)-\frac{R^{4}}{a^{4}}\left(\frac{z}{b}-\frac{z^{2}}{b^{2}}\right)\right.\nonumber\\
 \left.  +\ {\textstyle{\frac{1}{3}}}\left(\frac{z^{3}}{b^{3}}-1\right)+\frac{R^{4}}{a^{4}}\left(\frac{z^{2}}{b^{2}}-\frac{z^{3}}{b^{3}}\right)\right].\ \ \ \ \ 
\label{eq:13}
\end{eqnarray}
This then determines the components of the electric field ($\bE=\grad \Phi_{nid}$) everywhere in $D$ as
 \begin{equation}
E_{R}=\frac{\p\Phi_{nid}}{\p R}=\frac{2\Phi_{nid0}R^{3}}{a^{4}}\left(\frac{z}{b}-\frac{2z^{2}}{b^{2}}+\frac{z^{3}}{b^{3}}\right), \nonumber
\end{equation}
 \begin{equation}
E_{z}=\frac{\p\Phi_{nid}}{\p z}=\frac{\Phi_{nid0}}{2b}\left(1+\frac{R^{4}}{a^{4}}-\frac{z^{2}}{b^{2}}-\frac{4zR^{4}}{ba^{4}}+\frac{3z^{2}R^{4}}{b^{2}a^{4}}\right).  \nonumber
\end{equation}

In order to find $\Phi_{nid} (R,z)$ in region (4) of Figure 6, we start with the values of $\Phi_{nid}$ at the point $Q(a,z_{s})$ on the side of the diffusion region (Figure 6) and then calculate $\Phi_{nid}$ at any point $P(R,z)$ that lies on the same field line in region (4) to the right of $Q$.  Thus, after putting $(R,z)$=$(a,z_{s})$ in the expression (\ref{eq:13}) for $\Phi$ that holds in the diffusion region, we obtain
\begin{equation}
\Phi_{nid}(a,z_{s})\equiv f(z_{s})=-\Phi_{nid0}\left[\frac{1}{3}-\frac{z_{s}}{b}+\frac{z_{s}^{2}}{b^{2}}-\frac{z_{s}^{3}}{b^{3}}\right].
\label{eq:14}
\end{equation}
Since ideal MHD holds in region (4), $\Phi_{nid} (R,z)$ is constant along the field line ($zR^{2}=z_{s}a^{2}$) joining $Q$ to $P$, and so the value of $\Phi_{nid}$ at $P$ is simply
\begin{eqnarray}
\Phi_{nid}(R,z)=f\left(\frac{zR^{2}}{a^{2}}\right)\ \ \ \ \ \ \ \ \ \ \ \ \ \ \ \ \ \ \ \ \ \ \ \ \ \ \ \ \ \ \ \ \ \ \ \    \nonumber\\
=-\Phi_{nid0}\left[\frac{1}{3}-\frac{z}{b}\frac{R^{2}}{a^{2}}+\frac{z^{2}}{b^{2}}\frac{R^{4}}{a^{4}}-\frac{z^{3}}{b^{3}}\frac{R^{6}}{a^{6}}\right].
\label{eq:15}
\end{eqnarray}
The solution for $z<0$ can be obtained in a similar manner by integrating from $z = -b$.

We may now make various deductions from the solution. The reconnection rate depends on the form of $\eta$ and is given in order of magnitude by 
\begin{equation}
\int E_{\parallel}\ ds\ \sim\ 2E_{0}\ b,\nonumber
\end{equation}
where $E_{0}$ is the electric field at the centre of the diffusion region and $2 b$ is the dimension of the diffusion region along the magnetic field direction.  In our example, $E_{0}=E_{z}(0,0,0)=\Phi_{nid0}/(2b)=\eta j_{0}$, where $j_{0}={\bar j_{0}}B_{0}/(\mu L_{0})$ is the value of the current at the origin, and along the spine (\ref{eq:13}) implies that
\begin{equation}
E_{z}(0,0,z)=\frac{\Phi_{nid0}}{2b}\left(1-\frac{z^{2}}{b^{2}}\right),\nonumber
\end{equation}
and so the reconnection rate becomes more accurately
\begin{equation}
\int_{-b}^{b} E_{z}(0,0,z)\ dz\ = \ {\textstyle{\frac{4}{3}}}\ E_{0}\ b = \  {\textstyle{\frac{2}{3}}}\Phi_{nid0}.\label{eq:16}
\end{equation}

The other feature that we can deduce from the electric field components is the perpendicular plasma velocity given by Eq.~(\ref{eq:9}).
In particular, on the fan plane ($z=0$) inside $D$, $E_{R}=0$, $E_{z}=(\Phi_{nid0}/2b)(1+R^{4}/a^{4})$, $\eta j_{z}= (\Phi_{nid0}/2b)(1-R^{4}/a^{4})$ and $B_{R}=B_{0}R/L_{0}$ so that there is a rotational component given by
\begin{equation}
v_{\phi}=\frac{(E_{z}-\eta j_{z})B_{R}}{B^{2}}=v_{0}\frac{R^{3}}{a^{3}},\nonumber
\end{equation}
where $v_{0}=\Phi_{nid0}L_{0}/[baB_{0}(1+{\textstyle{\frac{1}{4}}}\ {\bar j_{0}}^{2})].$ The nature of the flow becomes clear if we subtract a component parallel to $\bB$ in order that $v_z=0$ (we are free to do this since the component of $\bv$ parallel to $\bB$ is arbitrary in the model). After doing this we find that  $v_R$ vanishes, leaving $\bv=(0, v_\phi, 0)$, i.e., the flow corresponds to a pure rotation (as in the solutions of Refs.~\cite{hornig03, pontin04a}).

\subsection{Extra Ideal Solution}\label{sec:3.2}
To the above pure diffusive solution any ideal solution may be added satisfying $\bE+\bv \x \bB={\bf 0}$ and $\curl \bE={\bf 0}$, for which the potential ($\Phi_{id}$) satisfies
\begin{equation}
\bB \. \grad \Phi_{id}=0.
\nonumber
\end{equation} 
Thus, once the functional form $\Phi_{id}(R_{0},\phi_{0})$ is chosen at the points $(R_{0},\phi_{0},b)$ on $z=b$, say, that form of $\Phi_{id}$ is constant along field lines given by the mapping (\ref{eq:11}). The resulting variation of $\Phi_{id}(R,\phi,z)$ throughout space is given by substituting for $R_{0}$ and $\phi_{0}$ from the inverse mapping (\ref{eq:12}). 

As an example, suppose
\begin{equation}
\Phi_{id}(R_{0},\phi_{0})=\Phi_{id0}\frac{R_{0}^{2}}{a^{2}}
\nonumber
\end{equation} 
on the plane  $z=b$.
Then throughout the volume we find
\begin{equation}
\Phi_{id}(R,\phi,z)=\Phi_{id0}\frac{R^{2}z}{a^{2}b},
\nonumber
\end{equation} 
which implies electric field components 
\begin{equation}
E_{R}=\frac{\Phi_{id0}}{a^{2}b}2Rz,\ \ \ \ \ \ E_{z}=\frac{\Phi_{id0}}{a^{2}b}R^{2}.
\nonumber
\end{equation} 
Then the plasma velocity components follow from $\bv_{\perp}=\bE\x\bB/B^{2}$ as
\begin{equation}
(v_{\perp R},v_{\perp \phi},v_{\perp z})=\frac{\Phi_{id0}L_{0}}{a^{2}bB_{0}}\frac{(-\half {\bar j_{0}} R^{3},R^{3}+4Rz^{2}, {\bar j_{0}}R^{2}z)}{(\alpha^{2}R^{2}+4z^{2})},
\label{eq:17}
\end{equation} 
where $\alpha^{2}=1+{\textstyle\frac{1}{4}}{\bar j_{0}}^{2}$.
In particular, we notice that the flow vanishes on the spine $R=0$, and that in the fan $z=0$ there is a rotational flow that linearly increases with distance $v_{\phi}(R,\phi,0)=-\Phi_{id0}L_{0}R/(a^{2}bB_{0}\alpha^{2})$.

The reconnection of field lines takes the form of a rotational slippage. 
Field lines entering the diffusion region have a flux velocity $\bw_{in}=-\grad \Phi_{in}\x \bB/B^{2}$,  while those that leave it have a flux velocity $\bw_{out}=-\grad \Phi_{out}\x \bB/B^{2}$. $\Phi_{in}$ is obtained by integrating along field lines that enter from the ideal region on one side, while  $\Phi_{out}$ is obtained by integrating backwards along field lines that leave from the other side. The rate of slippage between inward and outward flux bundles is given by $\Delta \bw= \bw_{out} - \bw_{in}$ and represents the rate of reconnection, which we have evaluated directly above in Equation (\ref{eq:16}). 
This reconnection rate, obtained by integrating $E_\parallel$ along the spine, measures the difference between the rates of flux transport across surface A and surface B in Fig.~\ref{fig:5}(c).  

Note that the extra ideal solution does not change the rate of relative slippage. However, it does allow for different external conditions, such as rotation above and below the diffusion region in the same or opposite senses. To see the effect of a non-rotational ideal flow see Ref.~\cite{pontin04a}. In the solution given above the physical quantities ${\bf E}$ and ${\bf v}$ are continuous but not differentiable at the boundary between regions (3) and (4). This is a sacrifice made for tractability and pedagogic purposes. For a solution with differentiable physical quantities, see \citet{pontin04a}.

In the above, the diffusion region was imposed to be a cylinder whose width ($a$) and height ($2b$) are parameters of the solution. The formation, in a self-consistent fashion, of such a cylindrical diffusion region was observed in the simulations described by  \citet{pontin07a}. In one of their simulations they imposed a twisting perturbation of the magnetic field in the vicinity of the fan plane. As the disturbance propagated inwards towards the null, it was dominated by a helical Alfv{\' e}nic wave -- travelling along the field lines and thus stretching out along the spine. The result was a tube of current focussed around the spine, giving a large aspect ratio to the diffusion region ($b \gg a$). During the process of torsional spine reconnection the narrowing and elongation of the current tube is likely  to continue until the rotational advection that twists the field and intensifies the current  is  balanced by the rotational slippage.

\begin{figure}
\centering
\includegraphics*[width=.29\textheight]{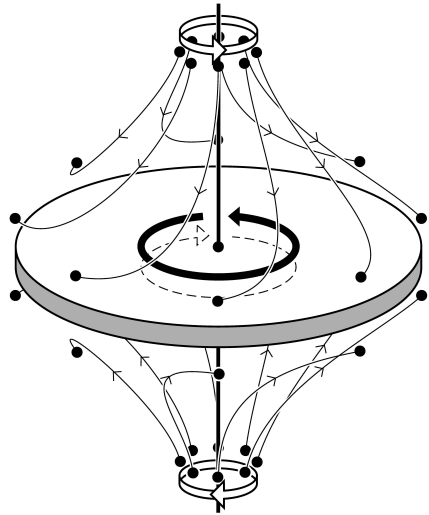}
\caption{A rotational motion of the spine (open arrows) driving torsional fan reconnection with a strong current in the fan and slippage of field lines (solid arrow).}
\label{fig:7}
\end{figure}

\section{Torsional fan reconnection}\label{sec:4}
Now suppose that we rotate the field lines near the spine in opposite directions above and below the fan.  Then a current builds up in the fan. Within the fan current sheet, field lines experience rotational slippage \cite{galsgaard03a,pontin07a}, in the opposite sense above and below the fan, in what we propose to term {\it torsional fan reconnection} (Figure \ref{fig:7}).  Again there is no flow across either spine or fan.

The counter-rotation (above and below the fan) of the region around the spine builds up a double-spiral structure near the null point, with a current that possesses two components: an axial component that reverses sign at the fan plane and a radial component.  A counter-rotating part to the diffusion velocity ($\eta j_{R} B_{z}$) is set up in the $\phi$-direction that reverses sign at the fan.

In order to model such reconnection, we consider what we term a {\it double-spiral\ null\ point} with field components
\begin{eqnarray}
(B_{R},B_{\phi},B_{z})=\frac{B_{0}}{L_{0}}\left(R,\ 2{\bar j_{0}}\frac{z^{2M+1} R^{N-1}}{b^{2M+N-1}},\ -2z\right)
\label{eq:18}
\end{eqnarray}
where $M$ and $N$ are positive integers and the corresponding current components are
\begin{eqnarray}
(j_{R},j_{\phi},j_{z})=\frac{2B_{0}{\bar j_{0}}}{\mu b^{2M+N-1}L_{0}}\ \ \ \ \ \ \ \ \ \ \ \ \ \ \ \ \ \ \ \ \ \ \ \ \ \ \ \ \ \  \nonumber \\
\left(-(2M+1)z^{2M}R^{N-1},\ 0,\ Nz^{2M+1}R^{N-2}\right).\nonumber
\end{eqnarray}
An alternative solution to the one presented below is outlined in the Appendix.

The field line equations for a mapping from an initial point $(R_{0},\phi_{0},b)$ to any other point $(R,\phi,z)$ are
\begin{eqnarray*}
R&=&R_{0}\ e^{S}, \ \ \ \ z=b\ e^{-2S},   \ \ \ \ \ \nonumber \\ 
\phi&=&\phi_{0}+\frac{2{\bar j_{0}}}{4M-N+4}\frac{R_{0}^{N-2}}{b^{N-2}}\left(1-e^{-(4M-N+4)S}\right),
\end{eqnarray*}
and the inverse mapping is 
\begin{eqnarray*}
R_{0}&=&R\ e^{-S}, \ \ \nonumber \\ 
\phi_{0}&=&\phi_{0}-\frac{2{\bar j_{0}}}{4M-N+4}\frac{R_{0}^{N-2}}{b^{N-2}}\left(1-e^{-(4M-N+4)S}\right),
\end{eqnarray*}
where $S=-{\half}\log(z/b)$.

Let us follow the approach of Section IIIA and calculate the pure non-ideal solution. We shall assume the diffusion region to be a disc of radius $a$ and height $2b$, with the same diagram as before (Figure 6), except that the diffusion region is now expected to be in the shape of a thin disc (with $b \ll a$) rather than a thin tube (with $b\gg a$).  Assuming, as before, that $\Phi(R,z)$ vanishes in regions (1) and (2), we evaluate it in region (3) by integrating from a point $T(R,b)$ on the top of the disc to a point $P(R,z)$ inside the diffusion region. After using Equation (\ref{eq:7}) and the mapping  and setting $\Phi_{e}=0$, the expression for the potential at $P(R,z)$ then becomes
\begin{eqnarray}
\Phi=-\Phi_{nid0}\int \frac{\eta}{\eta_{0}}\left( (2M+1)\frac{R_{0}^{N}}{b^{N}}e^{-(4M-N)S}\right. \nonumber \\
\left.+2N\frac{R_{0}^{N-2}}{b^{N-2}}e^{-(4M-N+6)S}\right) dS.
\label{eq:21}
\end{eqnarray}

We adopt the following general form for the magnetic diffusivity inside the diffusion region (D)
\begin{equation}
\eta=\eta_{0}\left(1-\frac{R^{m}}{a^{m}}\right)\left(1-\frac{z^{n}}{b^{n}}\right),
\nonumber
\end{equation}
which peaks at the null point and vanishes on the boundary of D when $m$ and $n$ are positive and $n$ is even. After substituting into (\ref{eq:21}) and using the mapping  and inverse mapping, we find the potential throughout the diffusion region. In particular, it transpires that an important constraint on the constants $M$, $N$, $m$ and $n$ is that  $E_{z}$  be finite and continuous at the fan plane.   As an example, one set of such constants that works is $M=2$, $N=6$, $m=4$ and $n=2$, for which 
\begin{eqnarray}
\Phi_{nid}(R,z)=-\Phi_{nid0}\left\{
\frac{z^2 R^4}{2b^6}
+ \frac{5z^{3}R^{6}}{3b^{9}}
-\frac{5z^{4}R^{6}}{2b^{10}} \right. \ \ \ \ \ \ \ \ \ \ \nonumber \\
\left. +\frac{5z^{5}R^{10}}{b^{11}a^{4}}
+\frac{5z^{6}R^{6}}{6b^{12}}
-\frac{5z^{6}R^{10}}{2b^{12}a^{4}}
+\frac{z^{8}R^{4}}{b^{12}}
-\frac{5z^{4}R^{10}}{2a^{4}b^{10}}  \right. \ \ \ \ \ \nonumber \\
\left. 
- \frac{3z^{6}R^{4}}{2b^{10}}
-\frac{3z^{4}R^{8}}{2b^{8}a^{4}}
- \frac{3z^{6}R^{8}}{b^{10}a^{4}} 
- \frac{3z^{8}R^{8}}{2b^{12}a^{4}}  
\right\}.\ \ \ \ \ 
\label{eq:22}
\end{eqnarray}
The corresponding components of electric field are
\begin{eqnarray}
E_{R}=\frac{\p\Phi_{nid}}{\p R}=-\frac{\Phi_{nid0}}{b}\left\{
\frac{2z^{2}R^{3}}{b^{5}} 
+\frac{10z^{3}R^{5}}{b^{8}} 
\right. \nonumber \\
-\frac{15z^{4}R^{5}}{b^{9}}
\left. +\frac{50z^{5}R^{9}}{b^{10}a^{4}}
+\frac{5z^{6}R^{5}}{b^{11}}
-\frac{25z^{6}R^{9}}{b^{11}a^{4}}
 \right. \nonumber \\
\left. +\frac{4z^{8}R^{3}}{b^{11}}
-\frac{25z^{4}R^{9}}{a^{4}b^{9}}
-\frac{6z^{6}R^{3}}{b^{9}}
\right. \nonumber \\
\left.
-\frac{12z^{4}R^{7}}{b^{7}a^{4}}
-\frac{24z^{6}R^{7}}{b^{9}a^{4}}
-\frac{12z^{8}R^{7}}{b^{11}a^{4}}\right\}, \nonumber
\end{eqnarray}
\begin{eqnarray}
E_{z}=\frac{\p\Phi_{nid}}{\p z}=
-\frac{\Phi_{nid0}}{b}\left\{
\frac{zR^{4}}{b^{5}}
+\frac{5z^{2}R^{6}}{b^{8}}
 -\frac{10z^{3}R^{6}}{b^{9}}\right. \nonumber \\
\left.+\frac{25z^{4}R^{10}}{b^{10}a^{4}}
+\frac{5z^{5}R^{6}}{b^{11}}
-\frac{15z^{5}R^{10}}{b^{11}a^{4}} \right. \nonumber \\
\left. +\frac{8z^{7}R^{4}}{b^{11}}
-\frac{10z^{3}R^{10}}{a^{4}b^{9}}
-\frac{9z^{5}R^{4}}{b^{9}}\right. \nonumber \\
\left.-\frac{6z^{3}R^{8}}{b^{7}a^{4}}
-\frac{18z^{5}R^{8}}{b^{9}a^{4}}
-\frac{12z^{7}R^{8}}{b^{11}a^{4}}\right\}, \nonumber
\end{eqnarray}

In order to find $\Phi(R,z)$ in region (4) of Figure 6, as before, we calculate its value at any point $Q(a,z_{s})$ where a field line leaves the diffusion region, and then project that value along that field line.  Thus, after putting $(R,z)=(a,z_{s})$ into Equation (\ref{eq:22}), we obtain 
\begin{eqnarray}
\Phi_{nid}(a,z_{s})\equiv f(z_{s})\ 
=-\Phi_{nid0}\left\{ 
\frac{z_{s}^2 a^4}{2b^6}
+\frac{5z_{s}^{3}a^{6}}{3b^{9}}
\right. \nonumber \\
\left. -\frac{5z_{s}^{4}a^{6}}{2b^{10}} 
+\frac{5z_{s}^{5}a^{6}}{b^{11}}
+\frac{5z_{s}^{6}a^{6}}{6b^{12}}
-\frac{5z_{s}^{6}a^{6}}{2b^{12}}
+\frac{z_{s}^{8}a^{4}}{b^{12}}  
\right. \nonumber \\
\left. 
-\frac{5z_{s}^{4}a^{6}}{2b^{10}} 
-\frac{3z_{s}^{6}a^{4}}{2b^{10}}
-\frac{3z_{s}^{4}a^{4}}{2b^{8}}
- \frac{3z_{s}^{6}a^{4}}{b^{10}} 
- \frac{3z_{s}^{8}a^{4}}{2b^{12}}  
\right\}.
\end{eqnarray}
Since ideal MHD holds in region (4), $\Phi_{nid} (R,z)$ is constant along the field line ($zR^{2}=z_{s}a^{2}$) joining $Q$ to $P$, and so the value of $\Phi_{nid}$ at $P$ is simply
\begin{eqnarray}
\Phi_{nid}(R,z)=f\left(\frac{zR^{2}}{a^{2}}\right) 
=-\Phi_{nid0}\left\{ 
\frac{z^2 R^4}{2b^6}
+\frac{5z^{3}R^{6}}{3b^{9}}
\right. \nonumber \\
\left. -\frac{5z^{4}R^{8}}{2a^{2}b^{10}} 
+\frac{5z^{5}R^{10}}{a^{4}b^{11}}
+\frac{5z^{6}R^{12}}{6a^{6}b^{12}}
-\frac{5z^{6}R^{12}}{2a^{6}b^{12}}
+\frac{z^{8}R^{16}}{a^{12}b^{12}}  
\right. \nonumber \\
\left. 
-\frac{5z^{4}R^{8}}{2a^{2}b^{10}} 
-\frac{3z^{6}R^{12}}{2a^{8}b^{10}}
-\frac{3z^{4}R^{8}}{2a^{4}b^{8}}
- \frac{3z^{6}R^{12}}{a^{8}b^{10}} 
- \frac{3z^{8}R^{16}}{2a^{12}b^{12}}  
\right\}.
\end{eqnarray}

The electric field  components vanish in both the spine and the fan but are strong just above and below the fan, which is where the reconnection of field lines occurs by rotational slippage in a similar fashion to torsional spine reconnection.  Near the spine and fan we have to lowest order in $R$ and $z$
\begin{equation} 
E_{R}=-\frac{\Phi_{nid0}}{b}\left(\frac{2z^{2}R^{3}}{b^{5}}\right),\ \ \  \ \ \ 
E_{z}=-\frac{\Phi_{nid0}}{b}\left(\frac{zR^{4}}{b^{5}}\right).\nonumber
\end{equation}  

The reconnection rate is the maximum value of ($\int E_{\parallel}\ ds$) along any field line ($R^{2}z=R_{0}^{2}b$), each of which enters the diffusion region from above at $T(R_{0},b)$ and leaves at $Q(a,z_{s})$. Along such field lines the integral is a function of $R_{0}/a$ and $b/a$, namely:
\begin{eqnarray*}
\int E_{\parallel}\ ds &=& \int \frac{\bE \. \bB}{B} ds = \int \frac{\bE \. \bB}{B_{R}} dR\ \ \ \ \ \ \ \ \ \ \ \ \ \ \ \ \ \ \ \ \ \ \ \ \ \ \ \ \ \ \nonumber \\
 &=& -\frac{\Phi_{nid0}}{b}\int 
 \frac{5z^{4}R^{5}}{b^{9}}
- \frac{5z^{6}R^{5}}{b^{11}}
+ \frac{5z^{6}R^{9}}{b^{11}a^{4}} \ \ \ \ \ \ \ \ \ \ \ \ \ \ \ \ \ \ \ \ \ \ \ \ \ \ \ \ \ \ \nonumber \\
&\ \ \ \ \ \ \ \ \ \ -&  \frac{12z^{8}R^{3}}{b^{11}}
- \frac{5z^{4}R^{9}}{b^{9}a^{4}}
+ \frac{12z^{6}R^{3}}{b^{9}} \nonumber \\
&\ \ \ \ \ \ \ \ \ \ +& \frac{12z^{6}R^{7}}{b^{9}a^{4}}
+ \frac{12z^{8}R^{7}}{b^{11}a^{4}} 
  dR \nonumber \\
 &=& \Phi_{nid0}
 \left[
 \frac{R_{0}^{4}}{2b^{4}}
 + \frac{5R_{0}^{6}}{3b^{6}}
 + \frac{9R_{0}^{8}}{2b^{8}}\frac{b^{4}}{a^{4}}
 \right.
 \ \ \ \ \ \ \ \ \ \ \ \ \ \  \nonumber \\ 
  &\ \ \ \ \ \ \ \ \ \  +&  \left. 
   \frac{5R_{0}^{10}}{b^{10}}\frac{b^{4}}{a^{4}}
  -\frac{5R_{0}^{8}}{b^{8}} \frac{b^{2}}{a^{2}}
  -\frac{5R_{0}^{12}}{3b^{12}} \frac{b^{6}}{a^{6}}
\right.
 \ \ \ \ \ \ \ \ \ \ \ \ \ \  \nonumber \\ 
  &\ \ \ \ \ \ \ \ \ \  -&  \left.   
     \frac{9R_{0}^{12}}{2b^{12}}\frac{b^{8}}{a^{8}}
     - \frac{R_{0}^{16}}{2b^{16}}\frac{b^{12}}{a^{12}} \right] . 
\end{eqnarray*}
When $R_{0}\sim a \gg b$ for a slender disc-shaped diffusion region, this reduces to
\begin{eqnarray*}
\int E_{\parallel}\ ds &=& -\frac{\Phi_{nid0} a^{4}}{2b^{4}}\left[
   \frac{R_{0}^{4}}{a^{4}} 
  + \frac{9R_{0}^{8}}{a^{8}} 
   - \frac{9R_{0}^{12}}{a^{12}} 
  - \frac{R_{0}^{16}}{a^{16}}  \right]. 
\end{eqnarray*}
If $a$ and $b$ are held fixed and $R_{0}$ is varied, the maximum value of this occurs at $R_{0}\approx 0.90 a$, giving a reconnection rate of
\begin{equation}
\left(\int E_{\parallel}\ ds\right)_{max} = 0.9 \ \Phi_{nid0}\ \frac{a^{4}}{b^{4}}.
\end{equation}

As for torsional spine reconnection, the reconnection rate is proportional to the potential $\Phi_{nid0}=2B_{0}b{\bar j_{0}}\eta_{0}/(\mu L_{0})$, but in this case, as well as being proportional to the current density ${\bar j_{0}}$ and diffusion region height ($b$), it also depends on its aspect ratio ($a/b$).

Again, as before, a wide range of ideal solutions ($\Phi_{id}$) may be added to the diffusive solution. Thus, if for example, $\Phi_{id}(R_{0},\phi_{0})=\Phi_{id0}R_{0}^{n}/a^{n}$ on the top ($z=b$) of the diffusion region, the fact that it remains constant along field lines ($R^{2}z=R_{0}^{2}b$) determines
\begin{equation}\Phi_{id}(R,\phi,z)=\Phi_{id0}\frac{R^{n}z^{n/2}}{a^{n}b^{n/2}},\nonumber\end{equation}from which the electric field components can be deduced.  

For instance, the case $n=4$ gives an electric field of\begin{equation}(E_{R},E_{\phi},E_{z})=\frac{2\Phi_{id0}}{a^{4}b^{2}}\left(2R^{3}z^{2},0,R^{4}z\right),\nonumber\end{equation}which implies a plasma velocity with components normal to the magnetic field of
\begin{eqnarray}
(v_{R},v_{\phi},v_{z})=\frac{1}{B^{2}}\left(-E_{z}B_{\phi},E_{z}B_{R}-E_{R}B_{z},E_{R}B_{\phi}\right)\ \ \ \ \ \ \ \ \ \ \nonumber \\
=\frac{\Phi_{id0}}{g(R,z)}\left(-4{\bar j_{0}}z^{6}R^{9},(2R^{5}z+8R^{3}z^{3})b^{9},8{\bar j_{0}}z^{7}R^{8}\right),
\nonumber
\end{eqnarray}where $g(R,z)\equiv [R^{2}+4R^{10}z^{10}{\bar j_{0}}^{2}/(b^{18})+4z^{2}]B_{0}a^{4}b^{11}/L_{0}$. In particular, it gives a rotational component ($v_{\phi}$) that is odd in $z$ and so represents the kind of counter-rotation that is typical of torsional fan reconnection.

\section{Spine-fan reconnection}\label{sec:5}
\begin{figure}
\centering
\includegraphics[height=.29\textheight]{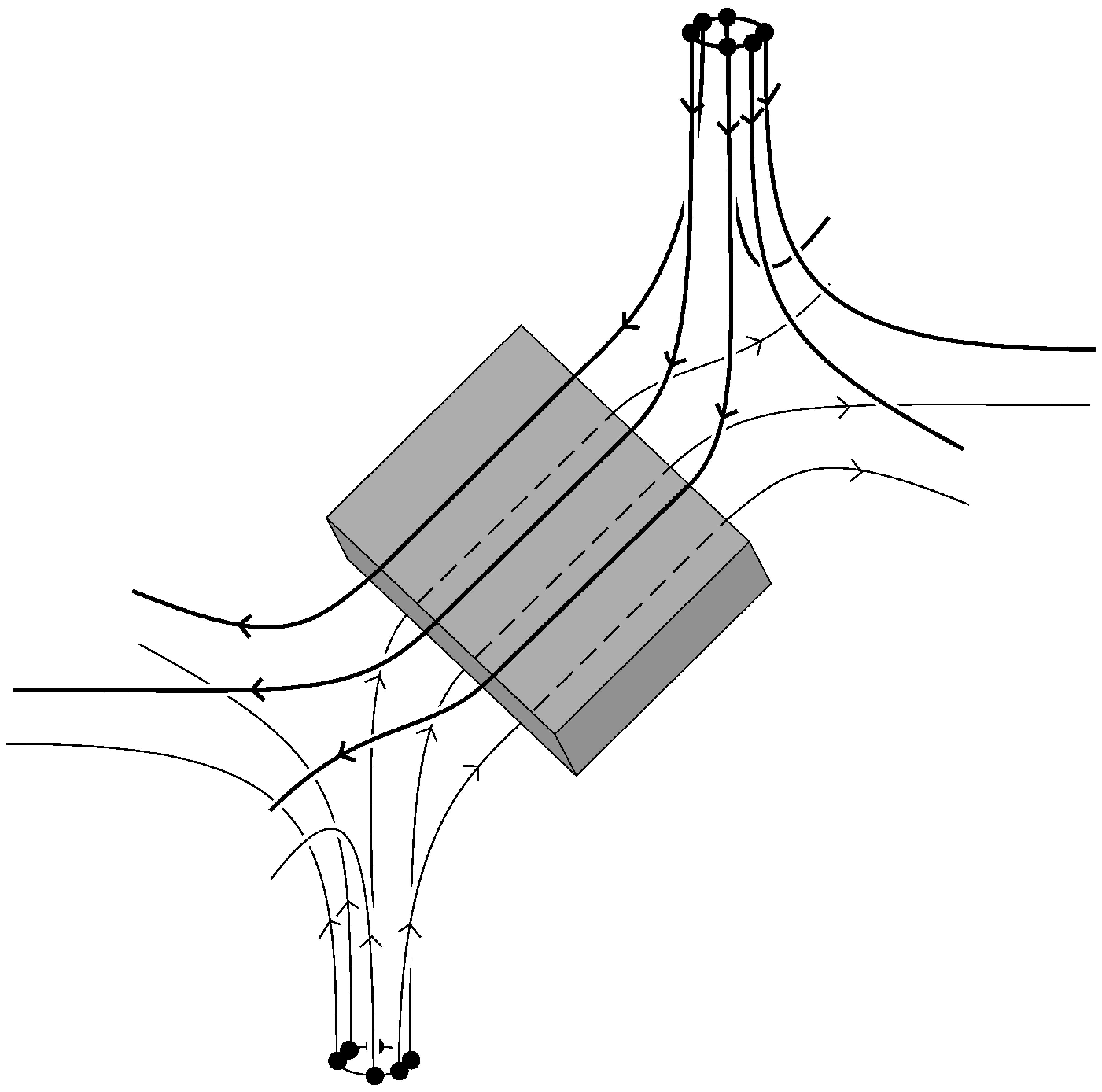}
\includegraphics[height=.29\textheight]{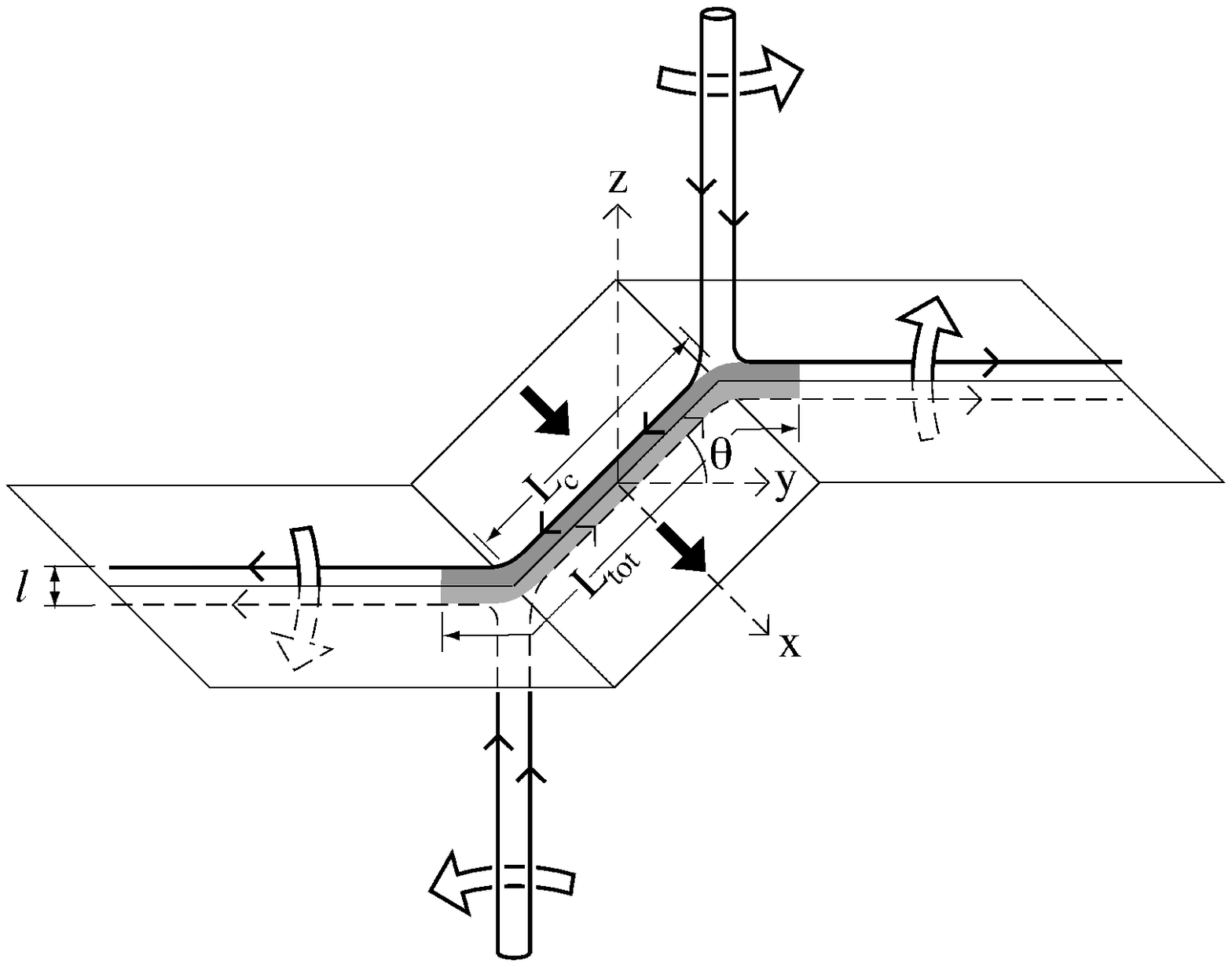}
\caption{(a) The magnetic structure of the field in spine-fan reconnection, showing the field lines and the (shaded) diffusion region. (b) The corresponding motion of flux across both the spine and fan (large light arrows). The current sheet is shaded (with the part below the fan having a lighter shading than the part above) and contains a current flowing in the $x$-direction (large dark arrows): its width is $l$, its total length $L_{tot}$ in the $yz$-plane, and its length $L_{c}$ common to spine and fan. }
\label{fig:8}
\end{figure}
In general, if the driving motions tend to shear a null point rather than rotate it, then the result will be {\it spine-fan reconnection}. A shear disturbance of either the spine or fan of the null will tend to make the null `collapse'. That is, the resulting Lorentz force acts to increase the displacement, just as at a 2D null \cite[see Ref.][]{pontin05a} and as at a separator \citep{galsgaard96b,galsgaard00b}. This collapse is opposed by the line-tying at the boundaries, and what occurs is that the shear distortion tends to focus in the weak field region in the vicinity of the null point, forming a localised current sheet \cite{pontin05a, pontin07b}. 

What distinguishes spine-fan reconnection from the other null point reconnection modes is that flux is transferred across both the spine and fan. Furthermore, the current concentration is in the form of a localised sheet that is inclined at an intermediate angle between the spine and fan -- indeed the current sheet contains part of both the spine and the fan (see Figures \ref{fig:4}b and \ref{fig:8}a). 
 As mentioned above, the reconnection rate for this mode of reconnection is obtained by integrating $E_\|$ along the fan field line with the maximum value of $\int E_{\|} ds$. By the symmetry of the simple models described herein, this is the field line parallel to the current orientation at the null (perpendicular to the applied shear). The reconnection rate thus obtained measures exactly the rate of flux transport in the ideal region across the fan separatrix surface. To illustrate the properties of this mode of null point reconnection, we describe here briefly the results of resistive MHD simulation runs (see \citet{pontin07b} for an initial description).

In the simulations, a shear velocity is prescribed on the (line-tied) $z$-boundaries, which advects the spine footpoints, see Fig.~\ref{fig:4}(a) (the results are qualitatively the same if the fan is distorted instead). The current sheet that forms in response to the shearing 
is localised in all three directions about the null. However, in the plane of the applied shear (perpendicular to the current orientation at the null) the magnetic field and current patterns have a similar appearance to a 2D X-point configuration. As one moves away from the null in the fan along the direction of current flow, the magnetic field strength parallel to the current (sometimes known as the `guide field') strengthens, while the current intensity weakens -- see Fig.~\ref{fig:8}.


The boundary shearing velocity is ramped up to a constant value ($v_0$) at which it is held until being ramped down to zero, at $t=\tau=3.6$ (space and time units in the code are such that an Alfv{\' e}n wave would travel one space unit in one time unit for uniform density and magnetic field $|\bB|=1, \rho=1$). The resistivity is uniform. Current focusses at the null during the period when the driving occurs, and when the driving ceases both the current and reconnection rate peak, after which the null gradually relaxes towards its original potential configuration. Under continuous driving, it is unclear whether a true steady state would  be set up, or whether the current sheet would continually grow in dimensions and intensity \cite[see Ref.][]{pontin07b}. This is an open question for future study. For the case of transient driving, the peak current and reconnection rate  increase linearly with the driving velocity. Here we examine more closely the sheet geometry, and its scaling with the driving velocity, and also investigate the scaling of this geometry, the peak current and peak reconnection rate with resistivity. 

As previously noted, the current sheet that forms in spine-fan reconnection is focussed at the null, locally spanning both the spine and fan. The sheet has a tendency to spread along the fan depending on the parameters chosen (spreading is enhanced by lowering $v_0$ or increasing the plasma-$\beta$). We examine four spatial measurements (see Figure \ref{fig:8}b) associated with the current sheet, focussing on the time when the current magnitude is a maximum and defining the boundary of the sheet to be an isosurface at 50\% of $|{\bf j}|_{max}$. The sheet {\it thickness} is $l$, the {\it length} $L_{tot}$ is the total extension in the $yz$-plane (normal to $\bj$), $L_c$ is the length of the `collapsed' section (within which the sheet contains both spine and fan), and the {\it width} $w$ is the extension of the sheet along the $x$-direction (parallel to $\bj$).

\begin{figure}[t]
\centering
\includegraphics[width=.5\textwidth]{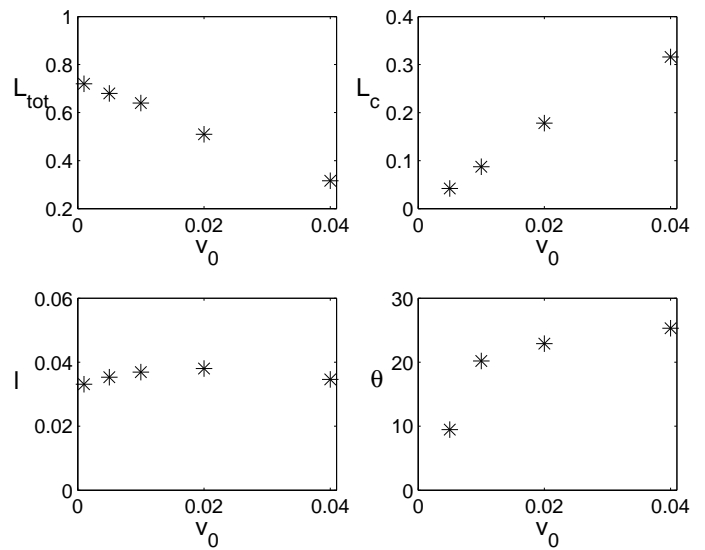}
\caption{Scaling with the driving velocity $v_0$ of (top left--bottom right) $L_{tot}$, $L_c$, $l$ and $\theta$ (see Figure \ref{fig:8}b for notation).}
\label{fig:9}
\end{figure}

The scaling of these dimensions with (peak) driving velocity $v_0$ is shown in Figure \ref{fig:9} (we fix $\eta=5\times10^{-4}$). The angle $\theta$ between the current sheet and the $z=0$ plane can be seen to increase as the driving velocity increases. This can be put down to the fact that the stronger driving creates a stronger Lorentz force---the force that is responsible for the collapse. 
As expected, $L_c$ increases as $v_0$ increases. This is a result of the fact that the spine footpoints are sheared further for larger $v_0$, and there exists in fact a close correspondence; $ (L_c \cos \theta)/2 \sim v_0 \tau $. In contrast to $L_c$, $L_{tot}$ shows a linear {\it decrease} with $v_0$ (as does $w$, see Ref.~\cite{pontin07b}), showing that as the collapse becomes stronger the distortion of the magnetic field focusses closer and closer around the null itself. The decline in $L_{tot}$ with increasing $v_0$ must of course cease once $L_{tot}=L_c$, as is the case for the strongest driving considered. Examining finally the sheet thickness $l$, any variation is within the error bars of our measurements, and moreover the resolution is not sufficient for firm conclusions to be drawn.

\begin{figure}[t]
\centering
\includegraphics*[width=.49\textwidth]{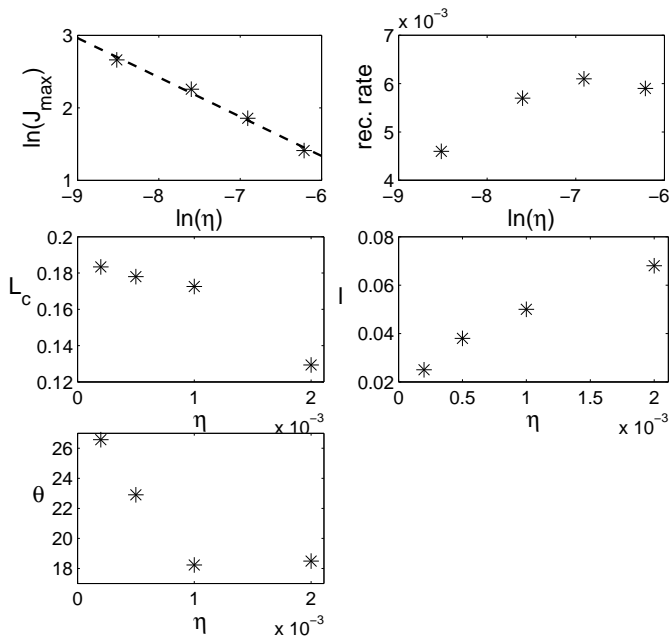}
\caption{Scaling with $\eta$ of (top left--bottom right) the peak current density, the peak reconnection rate, $L_c$, $l$ and $\theta$ (see Figure \ref{fig:8}b for notation).}
\label{fig:10}
\end{figure}

We turn now to consider the scaling of the current sheet with $\eta$, setting $v_0=0.02$, see Figure \ref{fig:10}. As $\eta$ decreases, $j_{max}$ increases, while the reconnection rate decreases. In both cases, with the limited data of this preliminary study, the proportionality appears to be somewhere between power law and logarithmic. That the run with the largest resistivity does not seem to fit the trend for the reconnection rate is likely to be because the current significantly dissipates before reaching the null itself due to the high resistivity ($\eta=0.002$).
Accompanying the increase in $j_{max}$ with $\eta$ is, as expected, a decrease in the thickness $l$. On the other hand, the overall dimensions of the sheet, $L_{tot}$ and $w$, seem to be unaffected by $\eta$, to within our measurement accuracy. Finally, as $\eta$ decreases and the current becomes more intense, the collapse becomes more pronounced as evidenced by increases in both $L_c$ and $\theta$. 

The relationships discussed briefly above certainly warrant further investigation with carefully designed, higher resolution simulations, as do the corresponding scalings for the continuously driven case.

\section{Conclusion}\label{sec:6}
We have here outlined a new categorisation of reconnection regimes at a 3D null point.  In place of the two previous types, namely, spine and fan reconnection, we suggest that three distinct generic modes of null point reconnection are likely to occur. The first two are caused by rotational motions, either of the fan or of the spine, leading to either {\it torsional spine reconnection} or {\it torsional fan reconnection}.  These involve slippage of field lines in either the spine or the fan, which is quite different from classical 2D reconnection, but does involve a change of magnetic connection of plasma elements.  

Even though pure spine or fan reconnection may occur in special situations (such as when $\nabla \cdot {\bf v}=0$ or  there are high-order currents), it is much more likely in practice that a hybrid type of reconnection takes place that we refer to as {\it spine-fan reconnection}. This is the most common form of reconnection that we expect to see in three dimensions at a null point, since it is a natural response to a shearing of the null point. It is most similar of all the 3D reconnection regimes to classical 2D reconnection and involves transfer of magnetic flux across both the spine and the fan. It possesses a diffusion region in the form of a current sheet that is inclined to the fan and spine and has current localised in both the spine and fan, focussed at the null.

In future, much remains to be determined about these new regimes of reconnection that have been observed in numerical experiments. One is the shape and dimensions of the diffusion regions and their relation to the driving velocity and the magnetic diffusivity.  Another key question is: what is the rate of reconnection at realistic plasma parameters, and is there a maximum value?  Since the analytical theory is so hard in three dimensions, progress is likely to be inspired by future carefully designed numerical experiments.

\section{Acknowledgments}
We are grateful to Guillaume Aulanier, Klaus Galsgaard, and our colleagues in the St Andrews and Dundee MHD Groups for stimulating discussions, especially Gunnar Hornig and Clare Parnell, and to the EU SOLAIRE network and UK Particle Physics and Astronomy Research Council for financial support. ERP is also grateful to Jiong Qiu, Dana Longcope and Dave McKenzie for inspiring suggestions in Bozeman where this work was completed.

\appendix

\section{Alternative Torsional Fan Solution}

Here we present another pure nonideal solution for torsional fan reconnection. It possesses a much simpler electric current, but the assumed form for the magnetic diffusivity has to be more complex and vanish on the fan.  We follow section \ref{sec:4} but consider a different form for the magnetic field of a double-spiral point, namely,
\begin{eqnarray}
(B_{R},B_{\phi},B_{z})=\frac{B_{0}}{L_{0}}\left(R,\ \frac{{\bar j_{0}}z R}{2b},\ -2z\right)
\label{eq:18}
\end{eqnarray}
and corresponding current components
\begin{eqnarray}
(j_{R},j_{\phi},j_{z})=\frac{B_{0}{\bar j_{0}}}{\mu bL_{0}}\left(-\half R,\ 0,\ z\right).\nonumber
\end{eqnarray}

The field line equations for a mapping from an initial point $(R_{0},\phi_{0},b)$ to any other point $(R,\phi,z)$ are
\begin{equation}
R=R_{0}\ e^{S}, \ \ \ \ z=b\ e^{-2S},   \ \ \ \ \phi=\phi_{0}+{\textstyle{\frac{1}{4}}}{\bar j_{0}}(1-e^{-2S}),
\label{eq:23}
\end{equation}
and the inverse mapping is 
\begin{equation}
R_{0}=R\ e^{-S}, \ \ \ \phi_{0}=\phi -{\textstyle{\frac{1}{4}}}{\bar j_{0}}(1-e^{-2S}),
\label{eq:24}
\end{equation}
where $S=-{\half}\log(z/b)$.

Assuming, as before, that $\Phi(R,z)$ vanishes in regions (1) and (2), we evaluate it in region (3) of Figure 6 by integrating from a point $T(R,b)$ on the top of the diffusion region to a point $P(R,z)$ inside the diffusion region. After using Equation (\ref{eq:7}) and the mapping (\ref{eq:23}) and setting $\Phi_{e}=0$, the expression for the potential at $P(R,z)$ then becomes
\begin{equation}
\Phi=-\Phi_{nid0}\int \frac{\eta}{4b^{2}\eta_{0}}(4b^{2}e^{-4S}+R_{0}^{2}e^{2S})dS.
\label{eq:25}
\end{equation}

As an example, we adopt the following form for the magnetic diffusivity inside the diffusion region (D)
\begin{equation}
\eta=\eta_{0}\left(1-\frac{R^{m}}{a^{m}}\right)\frac{z^{4}}{b^{4}}\left(1-\frac{z^{n-4}}{b^{n-4}}\right),
\nonumber
\end{equation}
which peaks above and below the null point and vanishes on the boundary of D. We have also chosen it to vanish on the fan plane since $j_{z}$ vanishes there and in order to facilitate a closed form solution with continuous physical quantities. After substituting into (\ref{eq:25}) and using the mapping (\ref{eq:23}) and inverse mapping (\ref{eq:24}), we find the potential throughout the diffusion region as
\begin{eqnarray}
\Phi_{nid}(R,z)=-\Phi_{nid0}\left \{\frac{1}{12}-\frac{z^{6}}{12b^{6}}+\frac{z^{m/2}R^{m}}{(m-12)b^{m/2}a^{m}}  \right.\nonumber \\
\left. 
-\frac{z^{6}R^{m}}{(m-12)b^{6}a^{m}}-\frac{1}{2n+4}+\frac{z^{n+2}}{(2n+4)b^{n+2}} \right. \nonumber \\
\left. 
- \frac{z^{m/2}R^{m}}{(m-2n-4)b^{m/2}a^{m}} +\frac{z^{n+2}R^{m}}{(m-2n-4)b^{n+2}a^{m}}  +\frac{zR^{2}}{24b^{3}} \right. \nonumber \\
 \left. -\frac{z^{4}R^{2}}{24b^{6}}+\frac{z^{m/2+1}R^{m+2}}{4(m-6)b^{m/2+3}a^{m}}  -\frac{z^{4}R^{m+2}}{4(m-6)b^{6}a^{m}} \right. \nonumber \\
 \left.  -\frac{zR^{2}}{(8n-8)b^{3}}+\frac{z^{n}R^{2}}{(8n-8)b^{n+2}}\right. \nonumber \\
 \left. -\frac{z^{m/2+1}R^{m+2}}{4(m-2n+2)b^{m/2+3}a^{m}}+\frac{z^{n}R^{m+2}}{4(m-2n+2)b^{n+2}a^{m}}
 \right \},
\label{eq:26}
\end{eqnarray}
and the corresponding components of electric field as
\begin{eqnarray}
E_{R}=\frac{\p\Phi_{nid}}{\p R}=-\frac{\Phi_{nid0}}{b}\left\{\frac{mz^{m/2}R^{m-1}}{(m-8)b^{m/2-1}a^{m}}\right. \nonumber \\
\left.-\frac{mz^{6}R^{m-1}}{(m-12)b^{5}a^{m}}-\frac{mz^{m/2}R^{m-1}}{(m-2n-4)b^{m/2-1}a^{m}}\right. \nonumber \\
\left.+\frac{mz^{n+2}R^{m-1}}{(m-2n-4)b^{n+1}a^{m}} +\frac{zR}{12b^{2}}-\frac{z^{4}R}{12b^{3}}\right. \nonumber \\
\left.+\frac{(m+2)z^{m/2+1}R^{m+1}}{4(m-6)b^{m/2+2}a^{m}}-\frac{(m+2)z^{4}R^{m+1}}{4(m-6)b^{5}a^{m}}\right. \nonumber \\
\left.-\frac{zR}{(4n-4)b^{2}}+\frac{z^{n}R}{(4n-4)b^{n+1}} -\frac{(m+2)z^{m/2+1}R^{m+1}}{4(m-2n+2)b^{m/2+2}a^{m}}\right. \nonumber \\
\left.+\frac{(m+2)z^{n}R^{m+1}}{4(m-2n+2)b^{n+1}a^{m}}
\right\}, \nonumber
\end{eqnarray}
\begin{eqnarray}
E_{z}=\frac{\p\Phi_{nid}}{\p z}=-\frac{\Phi_{nid0}}{b}\left\{-\frac{z^{5}}{2b^{5}}+\frac{m/2z^{m/2-1}R^{m}}{(m-12)b^{m/2-1}a^{m}}\right. \nonumber \\
\left.-\frac{6z^{5}R^{m}}{(m-12)b^{5}a^{m}}+\frac{z^{n+1}}{2b^{n+1}}\right. \nonumber \\
\left.-\frac{m/2z^{m/2-1}R^{m}}{(m-2n-4)b^{m/2-1}a^{m}}+\frac{(n+2)z^{n+1}R^{m}}{(m-2n-4)b^{n+1}a^{m}}\right. \nonumber \\
\left.+\frac{R^{2}}{24b^{2}}-\frac{z^{3}R^{2}}{6b^{5}}+\frac{(m/2+1)z^{m/2}R^{m+2}}{4(m-6)b^{m/2+2}a^{m}}\right. \nonumber \\
\left.-\frac{z^{3}R^{m+2}}{(m-6)b^{5}a^{m}}-\frac{R^{2}}{(8n-8)b^{2}}\right. \nonumber \\
\left.+\frac{nz^{n-1}R^{2}}{(8n-8)b^{n+1}} -\frac{(m/2+1)z^{m/2}R^{m+2}}{4(m-2n+2)b^{m/2+2}a^{m}}\right. \nonumber \\
\left.+\frac{nz^{n-1}R^{m+2}}{4(m-2n+2)b^{n+1}a^{m}}
\right\}, \nonumber
\end{eqnarray}

We note that the solutions in the lower half plane $z<0$ may be obtained simply by replacing $S=-{\half}\log(z/b)$ by $S=-{\half}\log(-z/b)$.  There is a term in $E_{z}$ that behaves like $R^{2}/b^{3}$ and so is usually discontinuous at the fan plane.  This discontinuity may, however, be avoided by setting $m=2$ and balancing it with the term in $z^{m/2}R^{m}$ when $a^{2}/b^{2}=12(n-1)(n+6)/[5(n-4)(n+1)]$.  As a simple example, let us consider $n=6$ and $a^{2}=72b^{2}/7$. Then, in order to find $\Phi(R,z)$ in region (4) of Figure 6, as before, we calculate its value at any point $Q(a,z_{s})$ where the field line leaves the diffusion region, and then project that value along the field line.  Thus, after putting $(R,z)=(a,z_{s})$ into Equation (\ref{eq:26}), we obtain 
\begin{eqnarray}
\Phi_{nid}(a,z_{s})=-\Phi_{nid0}\left[\frac{1}{48}+\frac{z_{s}}{7b}-\frac{171z_{s}^{2}}{280b^{2}}\right. \nonumber \\
\left.+\frac{3z_{s}^{4}}{14b^{4}}+\frac{29z_{s}^{6}}{120b^{6}}+\frac{z_{s}^{8}}{16b^{8}}\right].
\nonumber
\end{eqnarray}
As before ideal MHD holds in region (4), and so $\Phi_{nid} (R,z)$ can be calculated from the fact that it is constant along the field line ($zR^{2}=z_{s}a^{2}$) joining $Q$ to $P$. The electric field  components follow and can be used to calculate the reconnection rate and the flow velocity in the usual way.

Comparing the two forms of torsional fan solution, the advantage of the one presented in Section 4 is that the magnetic diffusivity peaks at the null point, but the slightly unwelcome feature is that the current density vanishes along the spine and fan.  By comparison, the solution in this appendix has a current that vanishes at the null point but nowhere else in the spine or fan, but its disadvantage is that we had to choose the diffusivity to vanish in the fan.


\end{document}